\documentclass[11pt]{article}
\usepackage{amssymb}
\usepackage{latexsym}
\usepackage{float}       
\usepackage{mathrsfs} 

\usepackage{graphicx} 
\usepackage{amsmath}
\usepackage{amssymb}  

\newcommand{\R}{{\mathbb R}}

\newcommand{\mI}{{\mathsf I}}

\newcommand{\mQ}{{\mathsf Q}}

\newcommand{\mT}{{\mathsf T}}

\newcommand{\mX}{{\mathsf X}}
\newcommand{\mS}{{\mathsf S}}

\newcommand{\mZ}{{\mathsf Z}}

\title{Mining the Mind: Linear Discriminant Analysis of MEG source reconstruction time series supports dynamic changes in deep brain regions during meditation sessions }
\author{D Calvetti$^1$ \and B Johnson$^1$ \and A Pascarella$^2$  \and F Pitolli$^3$ \and E Somersalo$^1$ \and B Vantaggi$^4$}
\date{$^1$ Case Western Reserve University \\ Department of Mathematics, Applied Mathematics and Statistics \\ 10900 Euclid Avenue, Cleveland, OH 44106 \\
$^2$ Istituto per le Applicazioni del Calcolo ``Mauro Picone'' - CNR \\ Via dei Taurini 19, 00185 Rome, Italy\\
$^3$ University of Rome ``La Sapienza''\\  Department of Basic and Applied Science for Engineering \\ Via Scarpa 16, 00161 Rome, Italy \\
$^4$ University of Rome ``La Sapienza''\\  Department MEMOTEF \\ Via del Castro Laurenziano 9, 00161 Rome, Italy
}
\begin{document}
 
\maketitle
\begin{abstract}
Meditation practices have been claimed to have a positive effect on the regulation of mood and emotion for quite some time by practitioners, and in recent times there has been a sustained effort to provide a more precise description of the changes induced by meditation on human brain.  Longitudinal studies have reported morphological changes in cortical thickness and volume in selected brain regions due to meditation practice, which is interpreted as evidence for effectiveness of it beyond the subjective self reporting. Evidence based on real time monitoring of meditating brain by functional imaging modalities such as MEG or EEG remains a challenge. In this article we consider MEG data collected during meditation sessions of experienced Buddhist monks practicing focused attention (Samatha) and open monitoring (Vipassana) meditation, contrasted by resting state with eyes closed. The MEG data is first mapped to time series of brain activity averaged over brain regions corresponding to a standard Destrieux brain atlas, and further by bootstrapping and spectral analysis to data matrices representing a random sample of power spectral densities over bandwidths corresponding to $\alpha$, $\beta$, $\gamma$, and $\theta$ bands in the spectral range. We demonstrate using linear discriminant analysis (LDA) that the samples corresponding to different meditative or resting states contain enough fingerprints of the brain state to allow a separation between different states, and we identify the brain regions that appear to contribute to the separation. Our findings suggest that cingulate cortex, insular cortex and some of the internal structures, most notably accumbens, caudate and putamen nuclei, thalamus and amygdalae stand out as separating regions, which seems to correlate well with earlier findings based on longitudinal studies.
\end{abstract}

\section{Introduction}

Meditation practices are widely recognized as  potentially powerful means for promoting physical and mental health through their acclaimed capability to reduce stress and anxiety \cite{Hofmann}, to increase concentration, cognitive performance and to improve self image \cite{Koole}. The benefits of meditation have been claimed for a long time by practitioners who have remarked on the positive effects of the practice, including, but not limited to, regulation of mood and emotions, and improved concentration power.  Since the effects of meditation practice are often based on self-reporting, questions have been raised about  bias towards positive effects \cite{Fox, Lutz2009}, making it hard to quantify and assess the clinical potential of meditation. In recent years, there has been a lot of research on the effects of meditation on brain \cite{Newberg}, however, the understanding is still far from complete.  The key elements across meditation practices are attention control, emotion regulation and self-awareness \cite{Tang, Grecucci}, thus motivating researchers to look for long-term or short-term changes in the associated brain regions. To determine whether the meditation benefits are only a placebo perception or the result of structural and functional changes in the brain, quantitative evidence to support the claims is needed, an approach advocated also by the Dalai Lama. 

In recent years, stress reducing lifestyle changes have been strongly advocated as an alternative to pharmacological intervention. As part of this movement, mindfulness meditation intervention has been considered as a potential aid from smoking cessation to control severe clinical depression. Concurrently,  the number of scientific publications investigating whether, and if so, how, meditation changes the brain, have increased at a steady rate. 

The literature on how meditation affects the brain can be subdivided into the contributions comparing the anatomy and functions of the brain of meditators versus control groups, and those focusing on the differences in brain functions during meditation sessions versus resting states. 
Cross-sectional studies look for differences between brains of meditators versus those of control groups, in particular, morphological and physiological differences in brain structure, including cerebral blood flow \cite{Newberg2}, cortical thickness and gray matter density \cite{Lazar}, and volume in brain regions \cite{Luders2012,Luders2013} that are believed to play a central role in meditation. In longitudinal studies, long term temporal changes due to meditation practice are monitored \cite{Gotnik1}. In both cases, confounding factors such as changes in lifestyle accompanying the decision to engage in extensive meditation practice  cannot be neglected, in particular since the test groups often involve professional meditators, who have dedicated years to this lifestyle changing practice. For these reasons, functional brain imaging modalities provide a potential, possibly less biased avenue towards a better understanding of the effects of meditation.

Functional brain imaging modalities such as fMRI, EEG and MEG provide potentially a way to follow the changes in brain activity patterns elicited by the meditation.  This article is contributing to the problem of identifying brain regions with most prominent changes in activity as the subject engages in meditative practices. The subjects considered in this study are professional meditators from the Theravada Buddhist tradition, and the data is collected during sessions corresponding to three brain states: The reference state with eyes closed, focused attention meditation (Samatha), and open monitoring akin to mindfulness meditation (Vipassana). In order to pinpoint the brain regions susceptible for activity changes when entering or exiting the meditative state, we first solve the MEG inverse problem, interpreting the biomagnetic signals in terms of the impressed currents in the brain, and summarize it by computing the aggregate activity over individual brain regions defined by a chosen brain atlas. The inversion algorithm is based on a hierarchical Bayesian approach, described and analyzed in \cite{IAS,Calvetti2}. Starting from the estimated aggregate activity in the brain regions, we generate large samples of the realizations of spectral densities over different frequency bands, corresponding to the $\theta$, $\alpha$, $\beta$, and $\gamma$ bands and carry out a spectral analysis. Applying this procedure to each brain state, we generate three large annotated data sets for each individual, and investigate them with data analysis tools. More specifically, our method of choice is linear discriminant analysis (LDA), a classical dimension reduction method that seeks to find the directions in the data space that best separate the annotated data sets. As the attributes of the data correspond to the different brain areas included in the atlas, the separating directions, in turn, can be interpreted in terms of brain regions, and the most significant components carry information about those that are more relevant for the separation of the states. Our analysis shows consistency between different meditation sessions of single individuals, and to some extent also between different individuals, indicating that the method is able to identify the key important brain regions involved in the meditative process. The findings are also in line with a number of meditation studies in the literature, as pointed out in the discussion of the results.

\section{Materials and methods}

The data consist of several hours of MEG recordings of professional meditators in the  Theravada Buddhist tradition with an average of over $15,750$ hours of meditation practice, recruited from the Santacittarama monastery in Italy. In addition to the regular daily two hours meditation practices, the monks take part in regular intensive meditation retreats, the practicing
comprising  two meditation styles: focused attention (Samatha) and open monitoring (Vipassana) meditation. In addition to data collected during the two meditation styles, reference data with no active meditation with eyes closed (rest) is provided. A detailed description of the data acquisition process, the selection of subjects, and the MEG device can be found in \cite{Marzetti}.

\subsection{From MEG data to activity vectors}

The raw MEG data was pre-processed by independent component analysis in order to
remove cardiac and eye movement artifacts. This
preprocessing step was performed within the open source package NeuroPycon
\cite{neuropycon} based on MNE Python routine \cite{mne}. The MRI
data of the subjects was segmented with Freesurfer \cite{Dale} and imported in
Brainstorm \cite{Tadel} to generate a subspace including both cortical surface
and substructure regions.
Subsequently, the time series data was processed by the iterative alternating sequential
(IAS) hierarchical Bayesian algorithm, which is described in detail in
\cite{IAS}, and analyzed further in \cite{Calvetti2}, resulting into a time
resolved dipole field time series $\{Q(t) = [\vec q_1(t), \ldots, \vec
q_M(t)]\}$ over the source space of cardinality $M$.  To reduce the
dimensionality of the brain signal, we generate an activity map over a selected
brain atlas by calculating the aggregate amplitude of the dipoles included in each
brain region (BR) in the atlas. Given an atlas, we define the {\em brain region activity level indicator} (BR-ALI) vector,
\begin{equation}
 {\bf a}(t) = \left[\begin{array}{c} a_1(t) \\ \vdots \\ a_n(t)\end{array}\right], \quad a_\ell(t) = {\rm sum}\{\|\vec q_j(t)\| \mid v_j\in{\mathcal R}_\ell\},
\end{equation}
where $\|\vec q_j\|$ denotes the intensity of the resolved dipole source at the source space vertex
$v_j$, and ${\mathcal R}_\ell$ is the $\ell$th parcel of source space vertices
corresponding to the $\ell$th BR in the atlas. In this work, we use the
Destrieux atlas for the cortical regions \cite{Destrieux2010}, consisting of 74
BRs per hemisphere, augmented by a parcellation of the internal structures as in
\cite{attal2013}, adding 8 BRs on both left and right, plus the brainstem. The total number of BRs in our analysis is $n = 165$. Observe that since no subsampling of the time series is performed, at the end of this process we have a time series of the BR-ALI vectors sampled with about one millisecond time resolution. The data acquisition protocol provides several copies of independently processed sequences of three different protocols (rest, Samatha, Vipassana), originating from different individuals. In this paper, we restrict the analysis to two meditators, referred to as Subject 1 and Subject 2.

\subsection{Bootstrapping, periodogram samples and power spectra}

The time dependent BR-ALI vectors ${\bf a}(t)$ corresponding to a selected protocol and subject are given at discrete times $t = t_k = k\Delta t$. The sampling frequency is denoted by $\nu_s = 1/\Delta t$. With $\Delta t \approx 1\, {\rm ms}$, the sampling frequency is $\nu_s \approx 1000\, {\rm Hz}$. In our calculations, we use the cosine transform, however, to conform with standard definitions, the complex exponential notation is used below.

\begin{figure}[ht!]
\centerline{
\includegraphics[width=12cm]{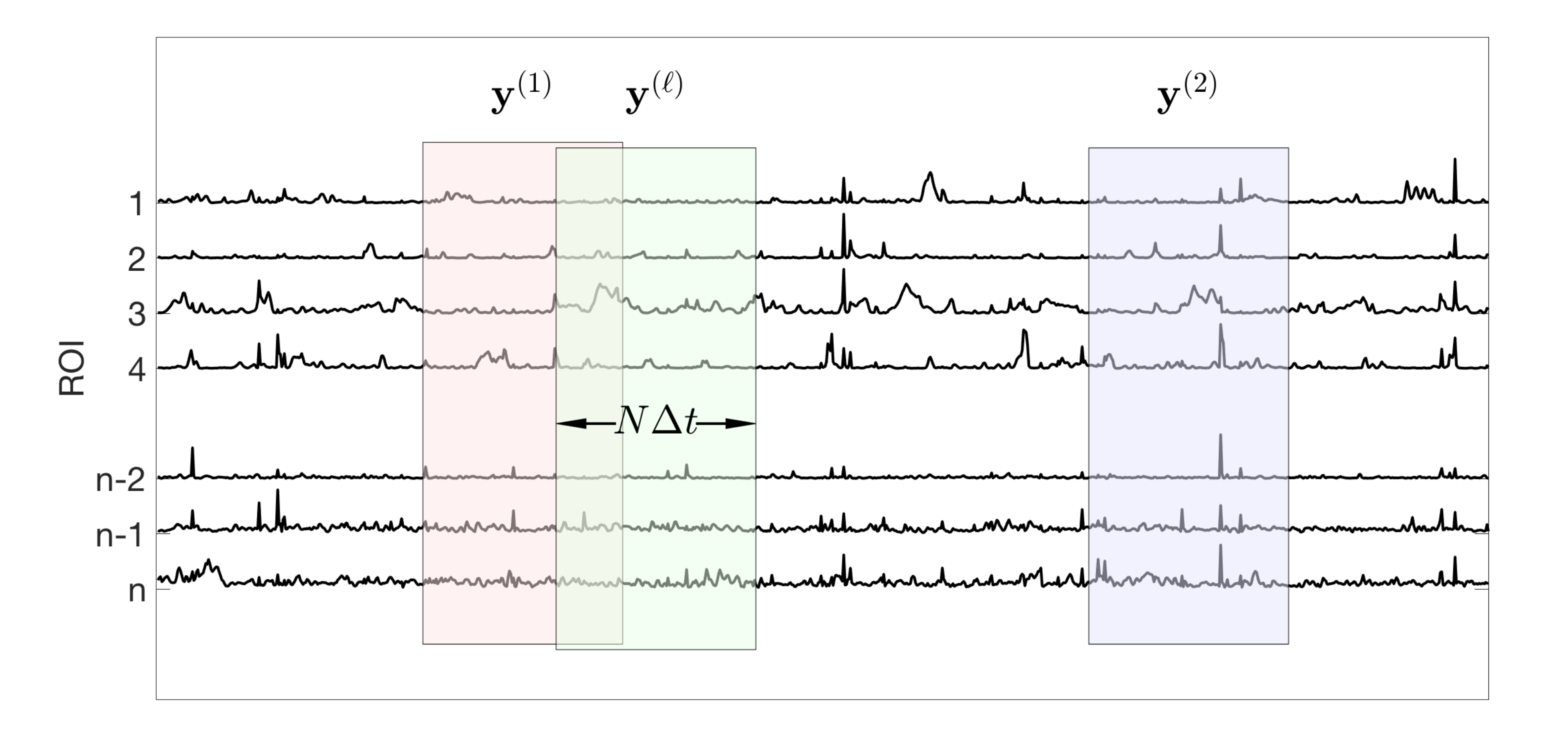}
}
\centerline{\includegraphics[height=8cm]{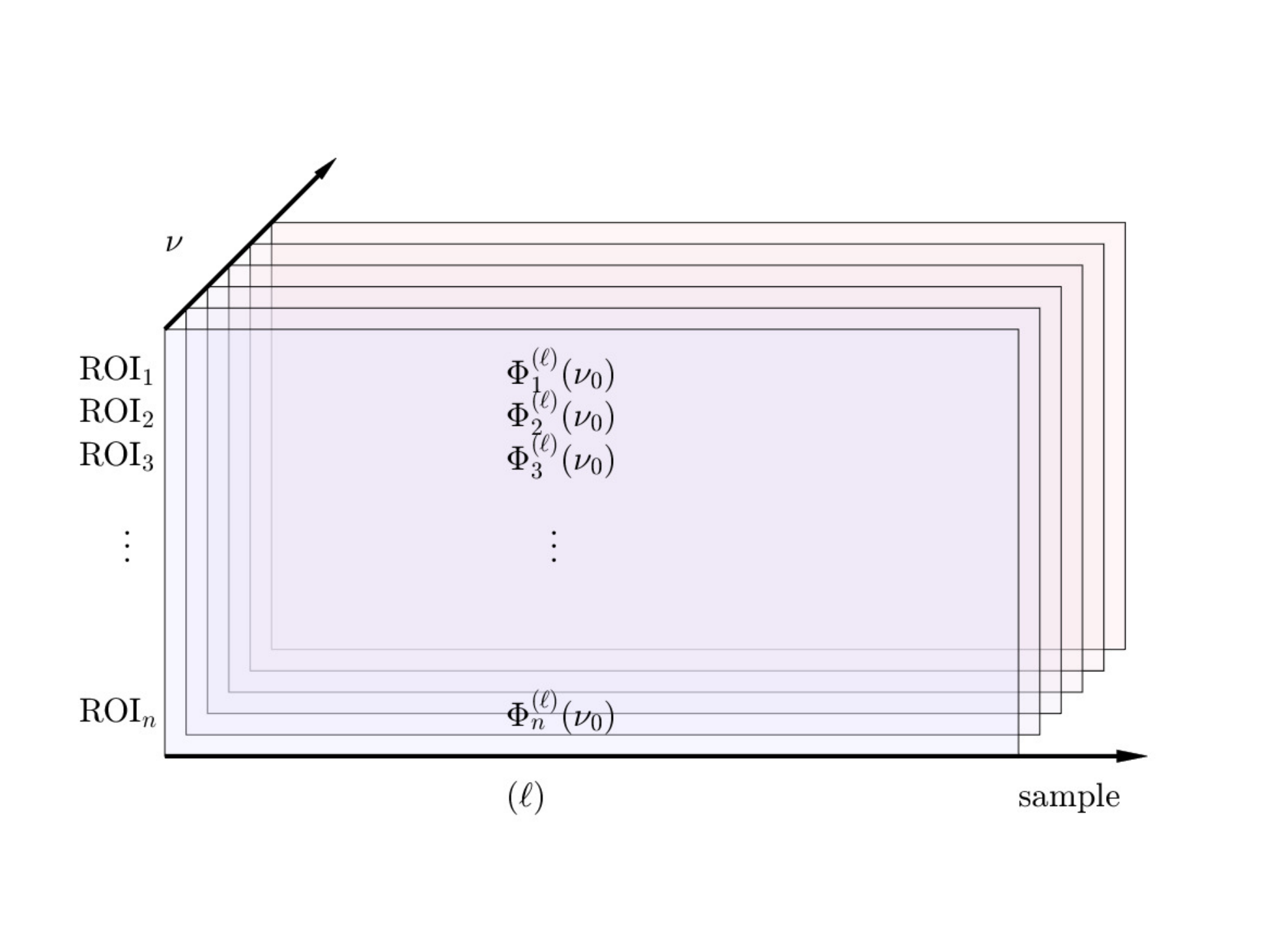}}

\caption{\label{fig:bootstrap}  Top: A schematic picture of the bootstrap process. The matrices ${\bf y}^{(\ell)}$ with entries $y^{(\ell)}_m(k)$ as defined in (\ref{sample signal}) comprise random cuts of fixed width $N\Delta t$  from the brain activity time series matrix. The sample vectors are Fourier transformed row-by-row, and the periodogram matrices are row-wise scaled squared amplitudes of the transforms, each row corresponding to a ROI, the columns referring to the frequencies. Bottom: Organization of the bootstrapped periodogram data into a three-dimensional array. Each array corresponds to one meditation protocol and one meditator. The power spectra are computed by averaging over the samples (dimension 2), and the spectral density sample matrices are computed by integrating (or in practice, summing) over a given spectral band (dimension 3).
}
\end{figure}

To analyze the BR-ALI vectors in the frequency domain, we start by recalling the definition of the power spectral density (PSD) of a scalar random signal $Y = (Y(0),Y(1),\dots,Y(N-1))$ of length $N$,
\begin{equation}
 {\rm PSD}(Y)(\nu) =  {\mathbb E}\left\{ \frac 1N \left| \sum_{j=0}^{N-1} Y(j) e^{-i2\pi\nu j\Delta t}\right|^2\right\} \approx 
 \frac 1 p \sum_{\ell=1}^p \Phi^{\ell}(\nu),
\end{equation}
where the expectation (${\mathbb E}$) is computed over periodograms of $p$ realizations of the signal \cite{StoicaMoses}, denoted by $y^{(\ell)} =(y^{(\ell)}(0),y^{(\ell)}(1),\ldots,y^{(\ell)}(N-1))$, $1\leq\ell\leq p$, and
\begin{equation}
 \Phi^{(\ell)} (\nu)  = \frac 1N \left| \sum_{j=0}^{N-1} y^{(\ell)}(j) e^{-i2\pi\nu j\Delta t}\right|^2, \quad 1\leq\ell\leq p.
\end{equation}
For our analysis, starting from the BR-ALI vectors with $n$ components corresponding to $n$ BRs, we define a sample of periodogram vectors using bootstrapping:
Given a sampling window of length $T_s = N\Delta t$, where $N$ is an integer, we denote by 
\begin{equation}\label{sample signal}
 {\bf y}^{(\ell)}(k) = \left[\begin{array}{c} y^{(\ell)}_1(k) \\ \vdots \\ y^{(\ell)}_n(k)\end{array}\right]  = {\bf a}(t^{(\ell)} + k \Delta t)\in\R^n, \quad k=0,1,\ldots,N-1,
\end{equation}
a random sample of length $T_s$ of the time series, where the initial value $t^{(\ell)}$, drawn randomly from the interval of observation, coincides with one of the discretization point. Figure~\ref{fig:bootstrap} elucidates the idea of the bootstrapping process. 

In the computations, we restrict the frequencies $\nu$ to a discrete grid,
\begin{equation}
 \nu = \nu_k = k/T_s, \quad 0\leq k\leq N-1,
\end{equation}
and, for each fixed frequency $\nu_k$, define the bootstrapped periodogram matrices of size $n\times p$,
\begin{equation}
 \Phi_m^{(\ell)}(\nu_k) = \frac 1N \left| \sum_{j=1}^N y^{(\ell)}_m(j) e^{-i2\pi\nu_k j\Delta t}\right|^2 =
  \frac 1N \left| \sum_{j=1}^N y^{(\ell)}_m(j) e^{-i2\pi j k/N}\right|^2, 
 \quad  \begin{array}{c} 1\leq \ell\leq p, \\ 1\leq m \leq n, \end{array}
\end{equation} 
or, in terms of the standard FFT of the bootstrapped sample, 
\begin{equation}
  \Phi_m^{(\ell)}(\nu_k) = \frac 1N\left|{\rm FFT}\big(y^{(\ell)}_m\big)_k\right|^2.
\end{equation}
We arrange the periodograms in a stack of random matrices as indicated in Figure~\ref{fig:bootstrap}, thus generating a three-dimensional array for each protocol and each meditator.

Observe that the DC component, $\nu_0 = 0$, corresponds to the average power in each ROI. 
The normalized PSDs of the ROIs, defined as
\begin{equation}\label{PSD scaled}
 \overline{{\rm PSD}}_m(\nu_k) = \frac{{\rm PSD}_m(\nu_k)}{{\rm PSD}_m(0)},
\end{equation}
represents the relative power density in the given frequency in each ROI.  

In this study, we are primarily interested in the distribution of the power among the spectral bands, $\alpha$- , $\beta$-, $\gamma$-, and $\theta$-bands, corresponding to different brain rhythms.
 For each brain rhythm, we identify a characteristic frequency band,
\begin{equation}
 I^i  = \left[\nu_{\rm min}^i,\nu_{\rm max}^i\right] ,\quad i \in\{\alpha,\beta,\gamma,\theta\},
\end{equation}
and define the sample matrix of bandwidth-specific integrated power vectors,
\begin{equation}
 \mX^{i} = \left[{\bf x}^{(i,1)},  {\bf x}^{(i,2)},\ldots, {\bf x}^{(i,p)} \right]\in\R^{n\times p},
\end{equation}
whose columns are sums of the scaled periodograms over the respective bandwidth, that is,
\begin{equation}
 {\bf x}^{(i,\ell)} = \left[\begin{array}{c}   x^{(i,\ell)}_1 \\ \vdots \\ x^{(i,\ell)}_n\end{array}\right], \quad
 x^{(i,\ell)}_m =   \frac{1}{\Phi^{(\ell)}_m(0)} \int_{\nu_{\rm min}^i}^{\nu_{\rm max}^i} \Phi^{(\ell)}_m(\nu) d\nu
 \approx \sum_{ \nu_k\in I^i} \frac{\Phi^{(\ell)}_m(\nu_k)}{\Phi^{(\ell)}_m(0)} \Delta\nu, \quad i\in \{\alpha,\beta,\gamma,\theta\}.
\end{equation}
In addition to the four spectral bands, we also consider the total power vectors, or the DC components, defined as
\begin{equation}
 {\bf x}^{({\rm DC},\ell)} = \left[\begin{array}{c}   x^{({\rm DC},\ell)}_1 \\ \vdots \\ x^{({\rm DC},\ell)}_n\end{array}\right], \quad
 x^{({\rm DC},\ell)}_m = \Phi^{(\ell)}_m(0),
\end{equation}
and define
\begin{equation}
 \mX^{\rm DC} = \left[ {\bf x}^{({\rm DC},1)}, {\bf x}^{({\rm DC},2)}, \ldots,  {\bf x}^{({\rm DC},p)}\right].
\end{equation}
The spectral bands used in our analysis are defined in Table~\ref{tab:bands}.
\begin{table}[ht]
\centerline{
\begin{tabular}{l|ccc}
 band & $\nu_{\rm min}$ & $\nu_{\rm max}$  & $\nu_{\rm peak}$\\
 \hline
 $\theta$ & $3.0\, (2.6) \,{\rm Hz}$ & $7.5\, (6.4)\,{\rm Hz}$   & $4.4 \,(3.7)\,{\rm Hz}$\\
 $\alpha$ & $7.5\, (6.4)\,{\rm Hz}$ & $12.0\, (10.2)\,{\rm Hz}$ & $8.9\,(7.7)\,{\rm Hz}$\\
 $\beta$ & $12.0\, (10.2)\,{\rm Hz}$ & $20.0\, (17.0) \,{\rm Hz}$  & $13.5\,(11.2)\,{\rm Hz}$ \\
 $\gamma$ & $25.0\, (21.3) \,{\rm Hz}$ & $40.0 \, (34.4)\,{\rm Hz}$ & - 
 \end{tabular}
}
\caption{\label{tab:bands} Brain activity bands used in the computation of the sample matrices $\mX^{i}$. The upper and lower bound of each band used in the calculation are $\nu_{\rm min}$ and $\nu_{\rm max}$, while  $\nu_{\rm peak}$ is the frequency at which the peak value occurs. In the $\gamma$ band, no unambiguous peak value was identified.  The numbers outside the parentheses are the frequencies used for Subject 1, and those in the parentheses refer to Subject 2. The adjustment of the bands is based on the observed offset of the positions of the $\theta$-, $\alpha$-, and $\beta$-peaks.}
\end{table}

The process described above is carried out separately for the data sequences corresponding to the three brain states (rest,Vipassana, Samatha). In this manner, using the independently processed copies of the data from the different meditators, we construct independent copies of bandwidth-specific data matrices  $\mX^i_{\rm rest}$, $\mX^i_{\rm Samatha}$, $\mX^i_{\rm Vipassana}$, where $i\in\{{\rm DC},\alpha,\beta,\gamma,\theta\}$.

\subsection{Activity patterns and LDA} 

For the time being, for each spectral band and each subject, we consider the data corresponding to the three different protocols: Eyes closed (rest), focussed attention (Samatha) and open monitoring or mindfulness (Vipassana) meditation, generate the scaled integrated periodogram samples and collect them in the form of $n\times p$ matrices  $\mX^{(1)} = \mX_{\rm rest}$, $\mX^{(2)} = \mX_{\rm Samatha}$, and $X^{(3)} = \mX_{\rm Vipassana}$. At this point we start addressing the following questions.
\begin{itemize}
\item[(i)] Is it possible to separate the three brain states on the basis of the periodogram matrices?
\item[(ii)] Which ROIs have more prominent roles in this separation?
\item[iii)] Are there significant differences between different spectral bands of the brain activity?
\item[(iv)] Is the separation signature for each subject consistent across temporally separated sessions?
\item[v)] Is the separation signature consistent across subjects?
\end{itemize}
The tool that we employ to answer these questions is linear discriminant
analysis, briefly summarized below. Historically, LDA was developed for separating two normally distributed samples \cite{Fisher}, with an underlying homoscedacity assumption. We point out that the version considered here is purely data driven and does not assume anything about the underlying distributions. We refer to standard data analysis and pattern recognition literature for details, see, e.g., \cite{Duda,CSdata}.

\subsubsection{LDA analysis}

Linear discriminant analysis is a standard dimension reduction technique, suitable to analyze the clustering of annotated data. Given an annotated set of vectors in $\R^n$, organized as columns of distinct matrices according to their annotation,
\begin{equation}
 \mX^{(1)}\in\R^{n\times p_1},\quad \mX^{(2)}\in\R^{n\times p_2},\quad \cdots,\mX^{(k)}\in\R^{n\times p_k},
\end{equation}
the LDA algorithm seeks a few vectors in $\R^n$ such that the orthogonal projections of each data set $\mX^{(\ell)}$ onto these directions appears as a compact cluster of points, with minimal  overlap of the projections of the different clusters. To give a precise meaning to this multi-objective optimization task, below we briefly review the ideas behind the LDA algorithm for finding the separating directions.

For each matrix $\mX^{(i)}$ whose columns are the vectors ${\bf x}^{(i,1)},\ldots, {\bf x}^{(i,p_i)}$ with identical annotation, referred to as a cluster, the corresponding $n\times n$ spread matrix  is defined as
\begin{equation}
  \mS^{(i)} = (\mX^{(i)} - \overline {\bf x}^{(i)}) (\mX^{(i)} - \overline {\bf x}^{(i)})^\mT, \quad \overline {\bf x}^{(i)} = \frac 1{p_i}\sum_{\ell=1}^{p_i} {\bf x}^{(i,\ell)},
\end{equation}
which is proportional to the empirical variance of the projected data.
Given a vector ${\bf q}\in\R^n$, the scalar quantity ${\bf q}^\mT \mS^{(i)}{\bf q}$ is, by definition, the spread of the cluster around the cluster mean $\overline {\bf x}^{(i)}$ in the direction determined by ${\bf q}$.  In statistical terms, the spread is proportional to the empirical variance of the projected data.
To find the directions in which all the clusters have a small spread, we use the {\em within-cluster spread} matrix as the sum of the individual spread matrices,
\begin{equation}
\mS_w = \sum_{i=1}^k \mS^{(i)},
\end{equation}
and observe that  ${\bf q}^\mT \mS_w{\bf q}$ quantifies the aggregate spread of the data in the direction ${\bf q}$ after each cluster has been centered around its mean.
Since we seek directions in which the clusters appear clearly apart from each others, to measure the separation we introduce the {\em between-cluster spread} matrix,
\begin{equation}
 \mS_b = \sum_{i = 1}^k p_i (\overline {\bf x}^{(i)} - \overline {\bf x})(\overline {\bf x}^{(\ell)} - \overline {\bf x})^\mT,
\end{equation}
where $\overline {\bf x}\in\R^n$ is the mean of the aggregate data regardless of the annotation. The between-cluster mean can be thought of as a spread matrix obtained by replacing each data vector in a cluster by the corresponding cluster mean, thus ignoring the spread within the individual clusters.

 Intuitively, the LDA algorithm seeks few directions such that the projected within-cluster spread in those directions is as small as possible, while the between-cluster spread is as large as possible. In other words, the LDA is looking for directions ${\bf q}\in\R^n$ for which the ratio
\begin{equation}
 H({\bf q}) = \frac{{\bf q}^\mT \mS_b {\bf q}}{{\bf q}^\mT \mS_w {\bf q}}
\end{equation}
is as large as possible. Observe that since the matrix $\mS_w$ could be singular, to avoid division by zero we consider instead the modified ratio 
\begin{equation}
 H_\delta ({\bf q}) = \frac{{\bf q}^\mT \mS_b {\bf q}}{{\bf q}^\mT(\mS_w +\delta \mI) {\bf q}},
\end{equation}
where $\delta>0$ is a small regularization parameter, and $\mI$ denotes the $n\times n$ unit matrix. 

The optimal projection direction is the vector ${\bf q}$ that maximizes the ratio $H_\delta({\bf q})$.  To find the optimal direction, we observe that $q$ solves of the generalized eigenvalue problem of finding a pair $({\bf q},\lambda)$ satisfying the equation
\begin{equation}
 \mS_b {\bf q} = \lambda(\mS_w + \delta \mI) {\bf q},
\end{equation}
Because all generalized eigenvalues $\lambda$ are real and non-negative, and the generalized eigenvector corresponding to the largest generalized eigenvalue is also the maximizer of $H_\delta({\bf q})$. Moreover, at most $k-1$ of the generalized eigenvalues are positive, thus there are at most $k-1$ separating directions. The LDA algorithm therefore seeks all the generalized eigenvectors that correspond to positive generalized eigenvalues. These vectors constitute the projection directions, and the dimensionality of the data can be reduced by representing the data points in terms the LDA components, namely the orthogonal projections onto the separating directions.  We point out that since, in general, the generalized eigenvectors are not mutually orthogonal, the LDA components may contain redundant information about the data. Orthogonalization of the LDA directions is possible, but usually leads to less clear cluster separation with little interpretative gain.

\subsubsection{Optimal window}

When computing the periodograms through bootstrapping, the window length $N$ in formula (\ref{sample signal})  plays an important role.
Asymptotically, when shrinking the window to a single time slice all interdependency between time slices will be lost: from the physiological point of view, it is not plausible that  snapshots over a window of one millisecond can reveal much of the current state of activity.  On the other hand, increasing the window length eventually will result into a loss of independency of the bootstrap samples. Intuitively, the window should be long enough to contain details of biological relevance, but short enough to not blur the details.
We therefore choose the window length based on separation power of the LDA. 

Given three data matrices $\mX^{(j)}\in\R^{n\times p}$, $1\leq j\leq 3$, the LDA will find up to two directions corresponding to positive generalized eigenvalues, denoted by
 ${\bf q}^{(1)},{\bf q}^{(2)}\in\R^n$. Thus, the LDA provides the means to represent the data in just two dimensions, with LDA components of each cluster matrix $\mX^{(j)}$ computed as
\begin{equation}
 \mZ^{(j)} = \mQ^\mT \mX^{(j)}\in\R^{2\times p}, \quad \mQ = \left[\begin{array}{cc} {\bf q}^{(1)} & {\bf q}^{(2)}\end{array}\right],
\end{equation}
where the vectors  $ {\bf q}^{(1)},  {\bf q}^{(2)} $ are of unit 2-norm. We define the {\em pairwise Bhattacharyya index} \cite{Bhattacharyya} to measure the mutual overlap as follows. 
Let $\Omega$ denote a rectangle containing the projected data points in the plane determined by the LDA separating vectors $ {\bf q}^{(1)},  {\bf q}^{(2)}$. 
We subdivide the rectangle $\Omega$  into $N$ bins, or pixels $\Omega_\ell$, $1\leq \ell \leq N$, and denote by ${\bf h}(\mX^{(j)}) \in\R^N$ the histogram vector of the projected data with components
\begin{equation}
 h_\ell(\mX^{(j)}) = \frac{\# (\mZ^{(j)}(1,k), \mZ^{(j)}(2,k))  \in\Omega_\ell}{p},
\end{equation}
or the number of projected data points in $\Omega_\ell$ normalized by the number of the data points. 
The pairwise Bhattacharyya index of the pair $(\mX^{(i)},\mX^{(j)})$ is 
\begin{equation}
 {\rm BI}(\mX^{(i)},\mX^{(j)}) = \sum_{\ell=1}^N \sqrt{h_\ell(\mX^{(i)}) h_\ell(\mX^{(j)})}.
\end{equation}
Observe that if we interpret the histograms as probability densities, the  quantity $1-{\rm BI}(X^{(i)},X^{(j)})$  is the Hellinger distance between the respective probability measures. The overlap measure can be extended to $c>2$ classes by introducing the {\em mean pairwise Bhattacharyya index} (MPBI), 
\begin{equation}
 {\rm MBPI}(\mX^{(1)},\ldots,\mX^{(c)}) = \frac{2}{c(c-1)}\sum_{i\neq j}^c  {\rm BI}(\mX^{(i)},\mX^{(j)}).
\end{equation}
In this study, we use $c=3$ for the three brain states.  Because the computation of the Bhattacharyya index depends on the binning of the rectangle $\Omega$, we need to specify a criterion for the discretization. We use the Bayesian optimal binning algorithm \cite{Knuth} to set the binning density.

\subsubsection{Separating vector analysis}

LDA produces few (in the current setting two) separating directions in the $n$-dimensional data space, which in the present case is 165, the number of brain regions in the selected parcellation. As the components of the separating vectors refer directly to the relative power over the bandwidth in a brain region, the separating directions may be interpreted as indicators of the brain regions most significant for the separation between the brain states. Note that in the face recognition literature, similar use of LDA has been proposed to identify differentiating features between sets of faces, and
the separating LDA directions are interpreted as face images, referred to as ``Fisher faces'' \cite{Belhumeur}.

Given the data matrices $\mX^{(j)}$, $j=1,2,3$ containing the data of the three different brain states of a single meditator over a prescribed frequency band, let ${\bf q}^{(1)},{\bf q}^{(2)}\in\R^n$ denote the two LDA vectors. Each component $q^{(k)}_{\ell}$,  $1 \leq \ell\leq n$ of the vector ${\bf q}^{(k)}$ refers to a brain region. 
If one of the $n$ components of the separator vectors vanishes, i.e.,  $q^{(1)}_{\ell} = q^{(2)}_\ell = 0$ for some $\ell$, this implies that the components $z_k^{(j)} = ({\bf q}^{(k)})^\mT \mX^{(j)}$ are insensitive  to the brain activity in the $\ell$th region. In that case, we may conclude that the $\ell$th brain region must play no role in the LDA separation, and therefore shows no particular specificity in the rest versus meditation practices. Conversely, if the $\ell$th component $q_{\ell}^{(k)}$ is large in one of the separator vectors, we may expect that the $\ell$th brain region has been identified as significant for the separation process. A similar reasoning is followed in the face recognition problem: A Fischer face reveals the facial features that are picked up by the LDA separation process, as opposed to generic features of little importance. Analogously, the separating vector significance hypothesis assumes that large components of the separating vectors identify brain regions that separate the meditative states from each other and from rest.

The separating vector analysis that considers only the brain regions corresponding to large components of the separating vectors, while appealing, may be subject to a fallacy for at least two reasons. First, LDA  is  seeking directions that not only have optimal separation power, but along which the clusters appear as compact as possible. Therefore, a brain region may be identified as relevant for its effectiveness in the compactification of the clusters. Second, since the absolute activity levels differ from each other, it is possible that a strongly separating brain region with high absolute activation level may not require a large coefficient $q_{j,\ell}$ to still play a role. Summarizing, it is safer to assume that vanishing or almost-vanishing components correspond to insignificant regions than claiming that a large component is necessarily significant.

To mitigate the risk of over-intepretation, we test the separating vector significance hypothesis by means of an exclusion principle. More specifically, we apply a {\em sequential filtering process} that removes brain regions corresponding to relatively small separating vector components and repeating the LDA analysis with the reduced data. If the cluster separation with the reduced data is still reasonably clear, this can be seen as an indication that the separating vector analysis is able to identify regions with separating power. The data reduction can be repeated several times, producing a nested set of index vectors, sequentially narrowing down the brain regions with separating power.

The LDA analysis and the subsequent separating vector analysis can be run also by restricting the data vectors to selected brain regions. Natural reduction schemes entail considering only the left or the right hemisphere, or only cortical regions or internal structures.

\section{Results}

The methodology described in the previous section is applied to data collected from two experienced meditators, referred to as ``Subject 1'' and ``Subject 2''. For both subjects, we select two sessions of each of the three protocols (rest, Samatha, and Vipassana), of the duration of approximately one minute, or 60\,000 time slices, and referred to  as ``Set A'' and ``Set B'' for both subjects. The sessions are temporally well separated from each other and preprocessed individually and independently to avoid correlations between preprocessing artifacts. For the description of the preprocessing, we refer to \cite{Marzetti,Calvetti2}.
Hence, the raw BR-ALI data comprise 6 data sets, for a total of  $60\,000$ vectors in $\R^{165}$ for each subject.

\begin{figure}
\centerline{\includegraphics[height = 10cm]{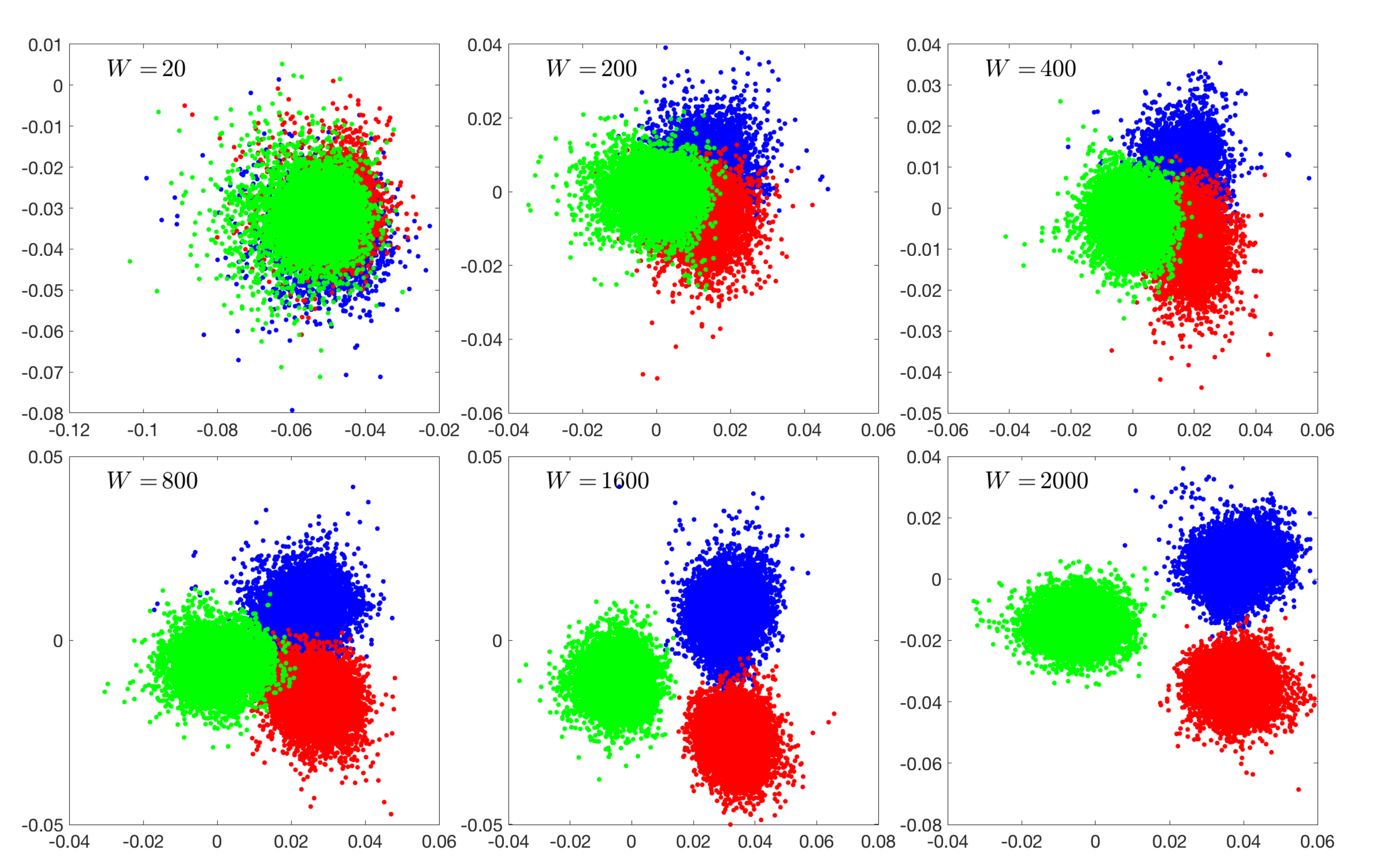}}
\caption{\label{fig:separation} Scatter plots of the LDA projections of the three meditation protocols of Subject 1 in the $\beta$ band, with different sampling window lengths used to compute the periodograms. The number of bootstrap samples is $10\,000$ for each protocol. The resting state (or eyes closed) is in blue, Samatha in red, and Vipassana in green. The images, in lexicographical order, correspond to window lengths 20, 200, 400, 800, 1200 and 2000.}
\end{figure}

\begin{figure}
\centerline{\includegraphics[width=8cm]{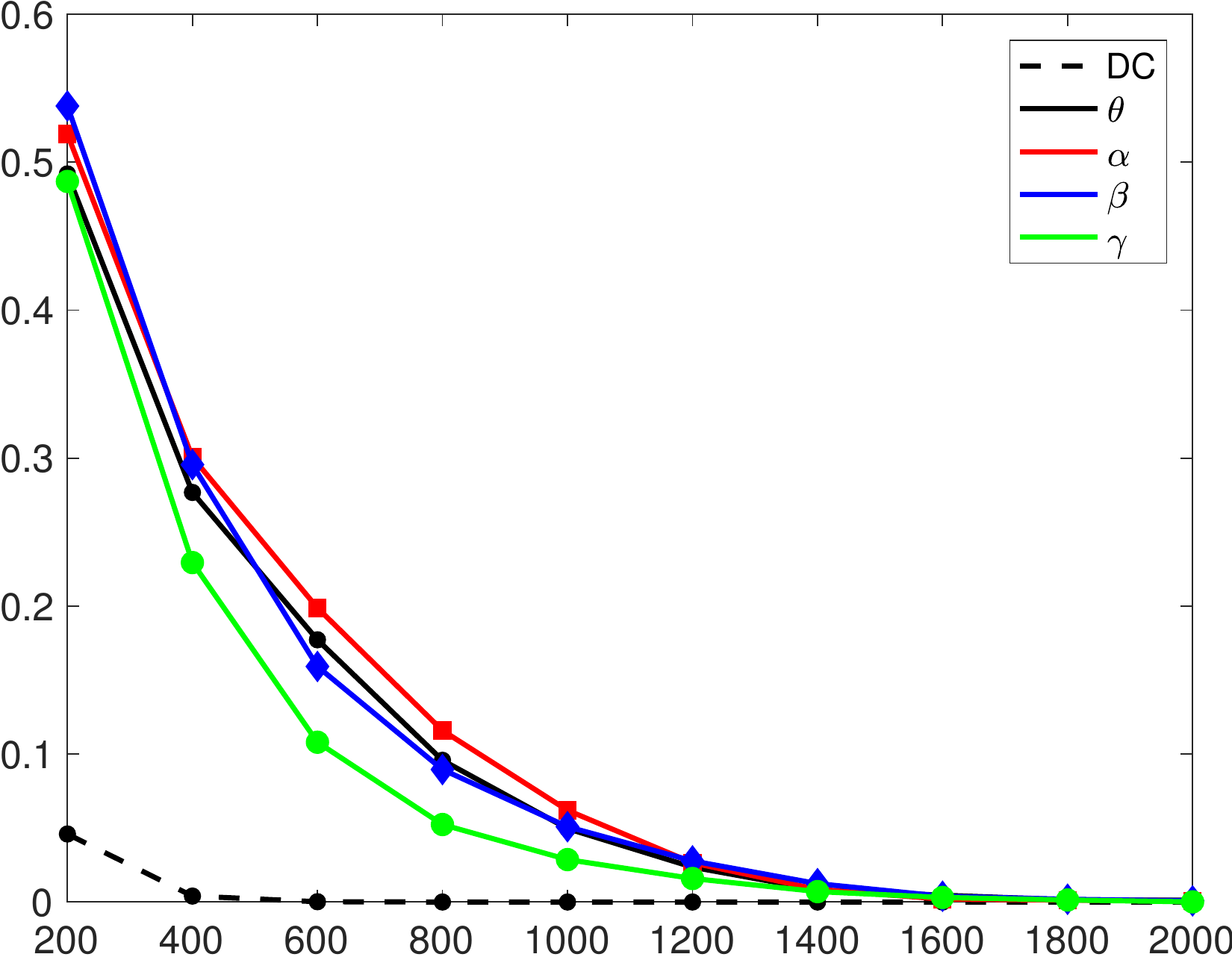}}
\caption{\label{fig:window}The ${\rm MBPI}$ overlap index computed with different bootstrap window lengths.}
\end{figure}

We start the LDA  by considering the optimal windowing in the bootstrap process. To analyze the cluster separation by LDA, we select one data sequence corresponding to each of the three brain states  (rest, Samatha, Vipassana) for Subject 1, and we compute  the matrices $\mX^{(j)}$, $j=1,2,3$ with different sampling window lengths $N$ (see formula (\ref{sample signal})). Each matrix has $p = 10\,000$ columns, and the sampling window length varies up to  2\,000. We then perform LDA for all selected spectral bands. Figure~\ref{fig:separation} shows the LDA projections of the $\beta$-band matrices of the three protocols for various window lengths, the corresponding scatter plots with other frequency bands being qualitatively very similar. As expected, short sampling windows do not allow a good separation, since for the method to recognize the characteristic spectral contents in the sample, and consequently to separate between the states, sufficiently long time series are needed. In Figure~\ref{fig:window}, the pairwise overlap index is plotted for window lengths from 200 to 2\,000 for each one of the spectral bands. We see that, with the exception of the DC band, a window length of almost 2\,000 is necessary to have complete separation of the three states. Based on this observation, we set the window length to 2\,000.  

\begin{figure}[ht]
\centerline{
\includegraphics[width=16cm]{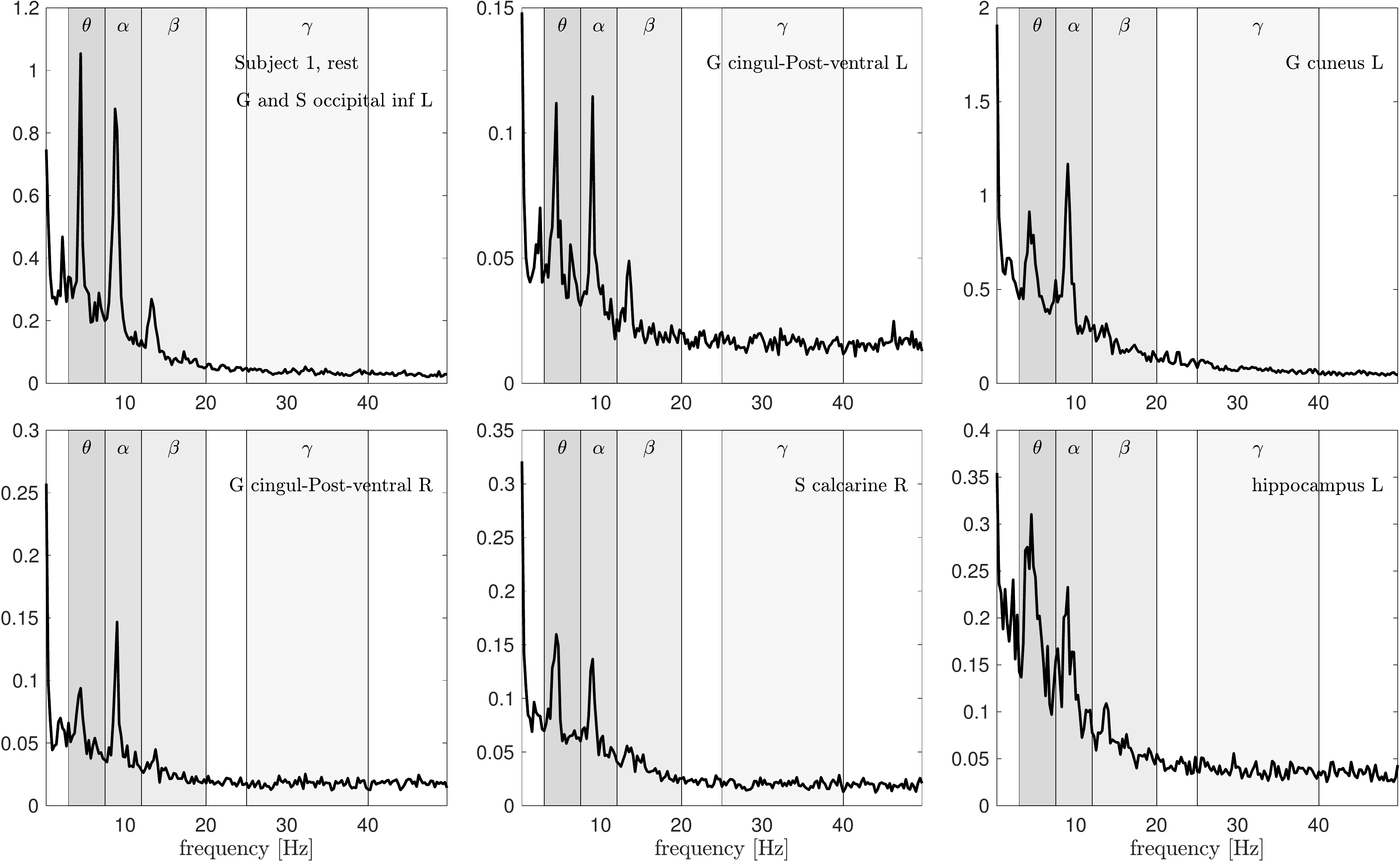}
}
\centerline{
\includegraphics[width=16cm]{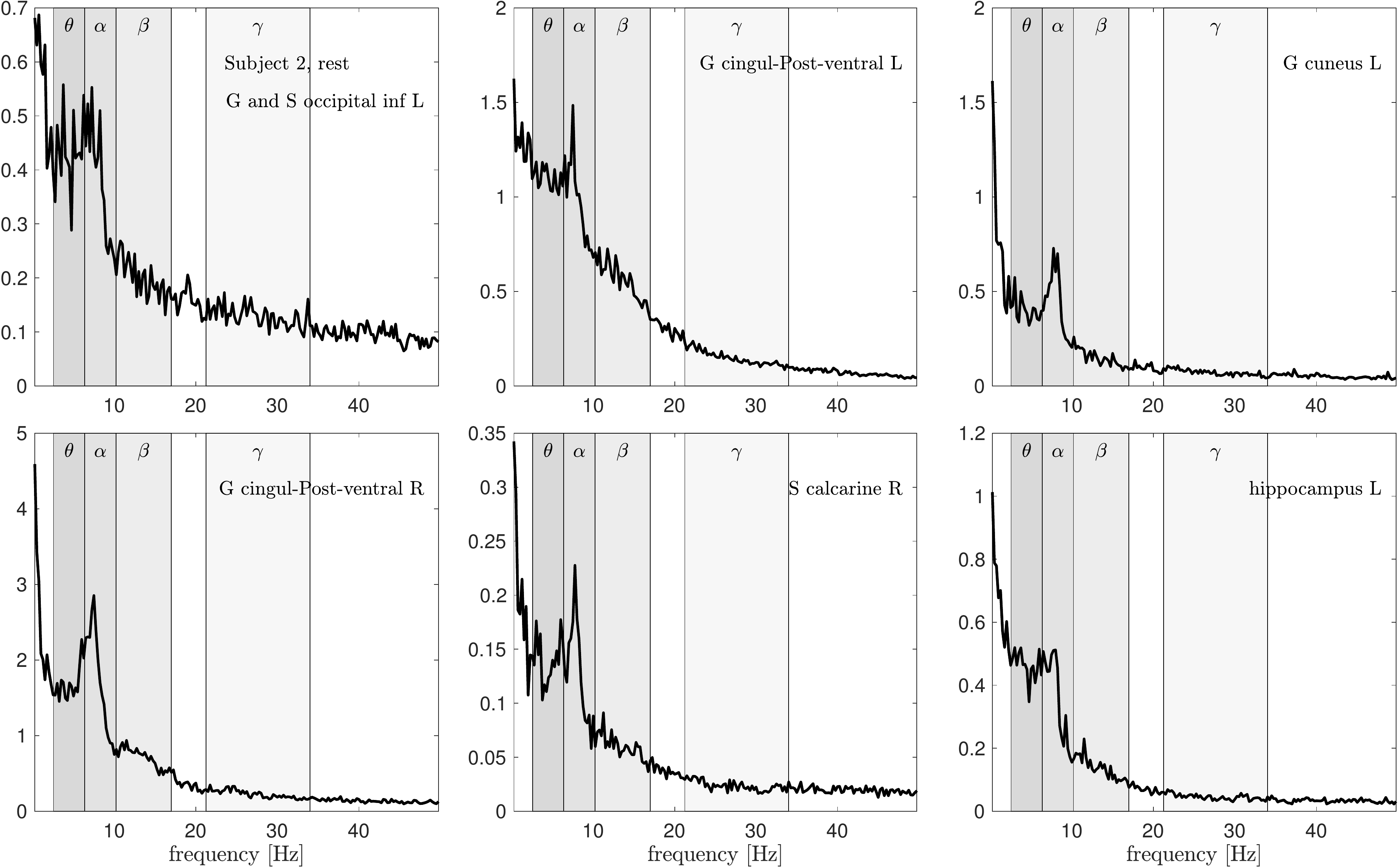}
}
\caption{\label{fig:PSD_1} Six scaled power spectral densities corresponding to eyes closed resting state of Subject 1 (two top rows) and Subject 2 (two bottom rows). The scaled PSD values have been scaled by a factor of 100. Observe that each PSD is scaled differently. The peak values for Subject 2 appear at a frequency lower than those of Subject 1, and the frequency bands a adjusted accordingly, see Table 1.}
\end{figure}

\begin{figure}[ht]
\centerline{
\includegraphics[width=16cm]{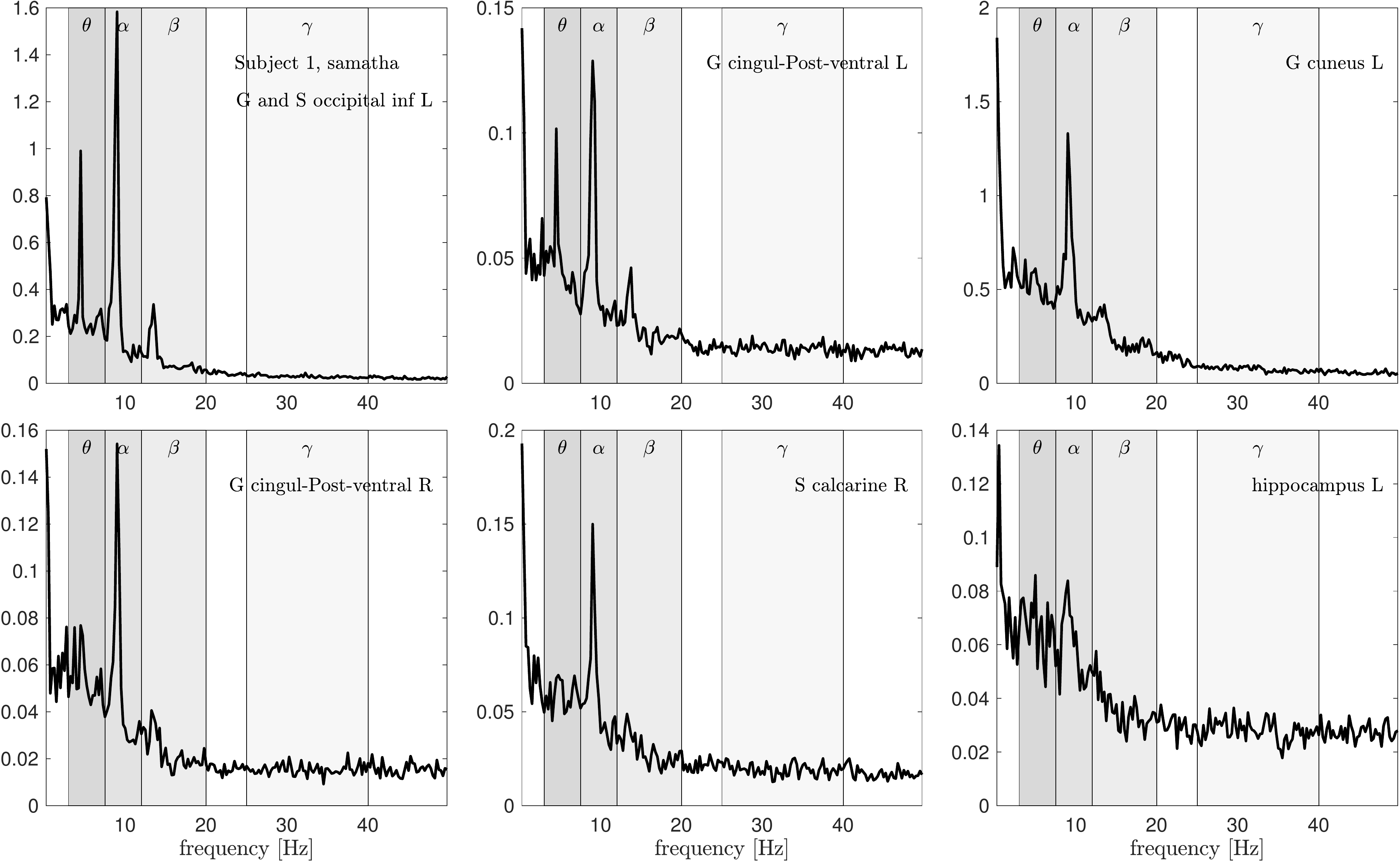}
}
\centerline{
\includegraphics[width=16cm]{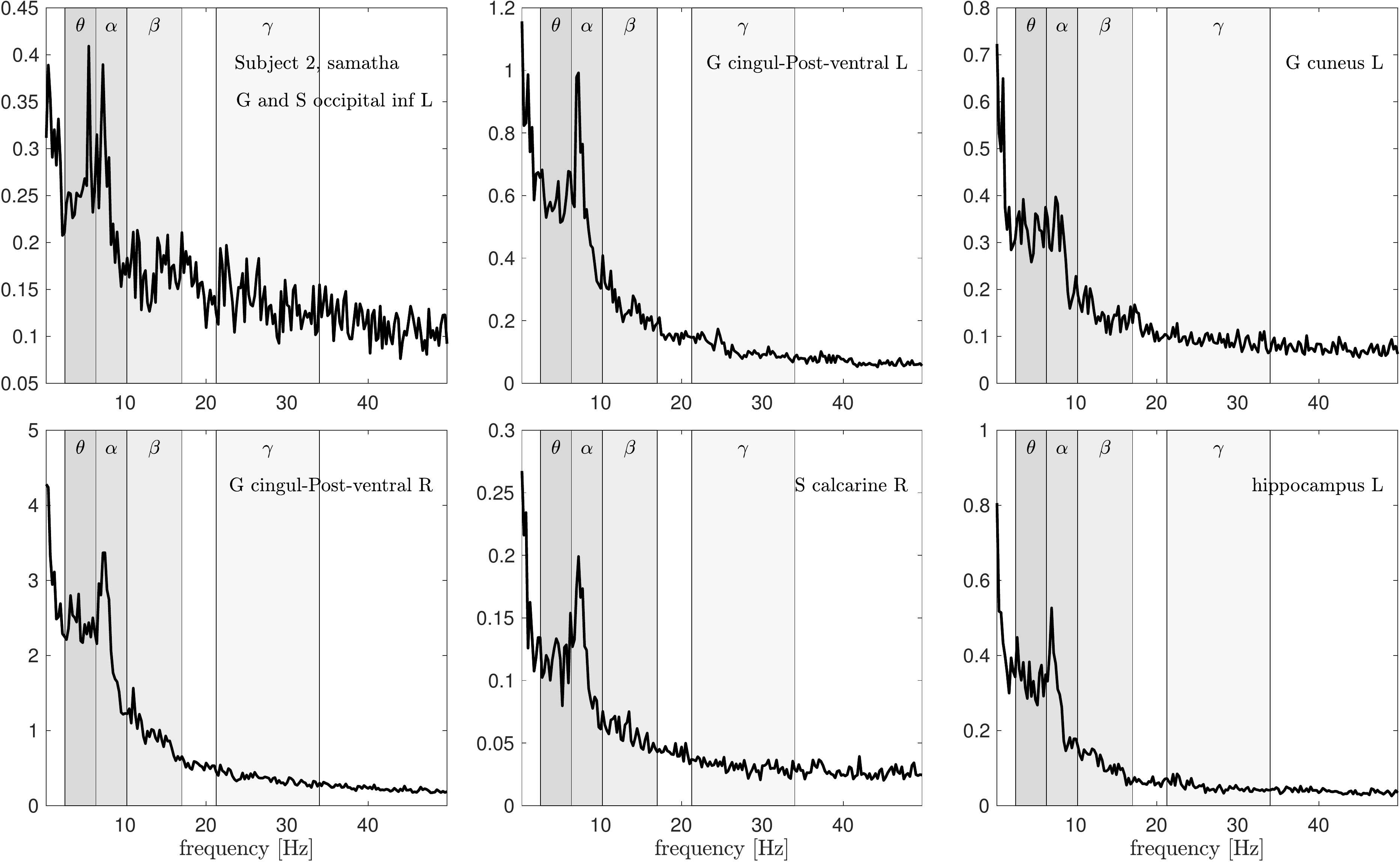}
}
\caption{\label{fig:PSD_2} Six scaled power spectral densities corresponding to focused attention (Samatha) meditation state of Subject 1 (two top rows) and Subject 2 (two bottom rows). The scaled PSD values have been scaled by a factor of 100. Observe that each PSD is scaled differently. The peak values for Subject 2 appear at a frequency lower than those of Subject 1, and the frequency bands are adjusted accordingly, see Table 1.}
\end{figure}

\begin{figure}[ht]
\centerline{
\includegraphics[width=16cm]{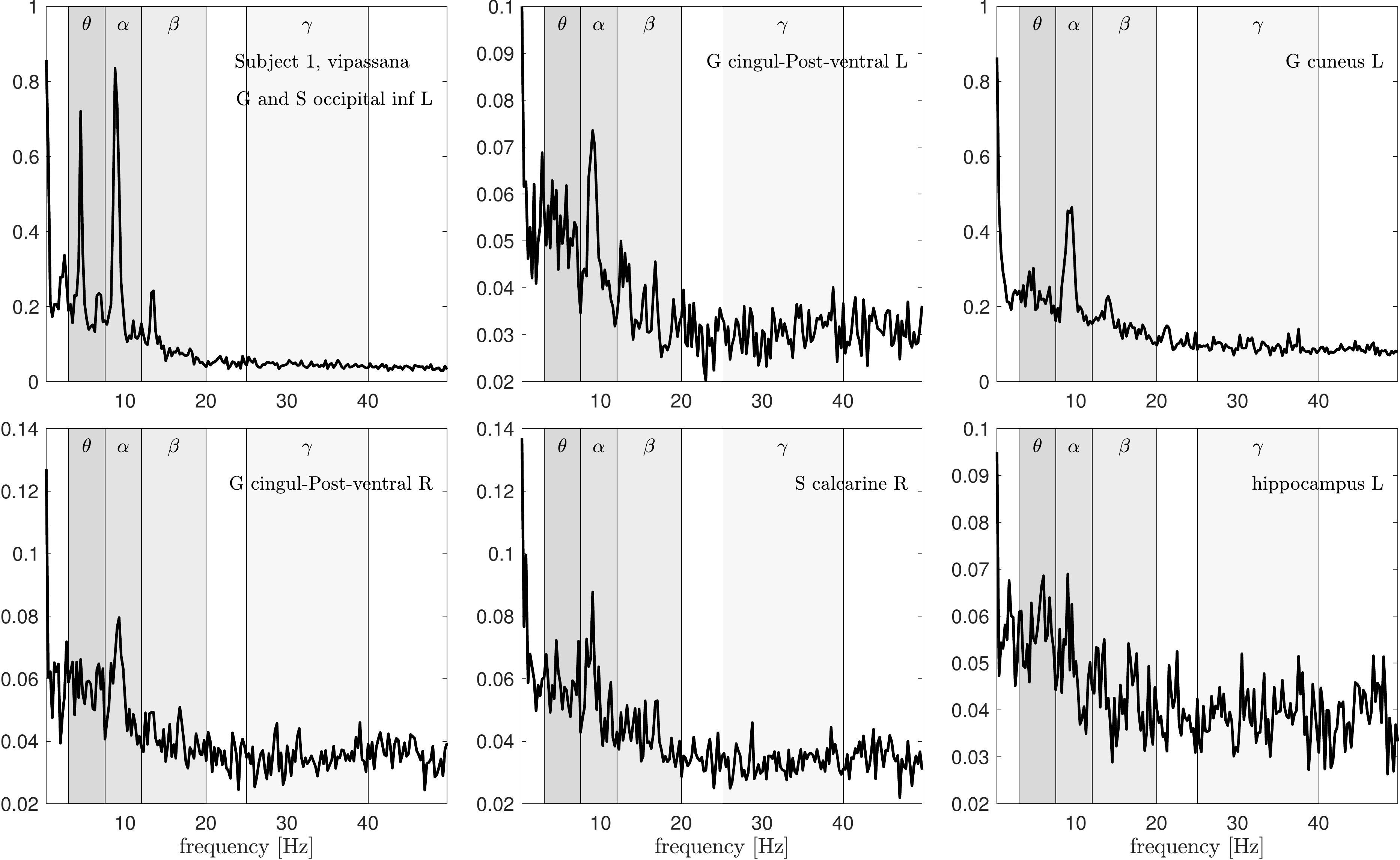}
}
\centerline{
\includegraphics[width=16cm]{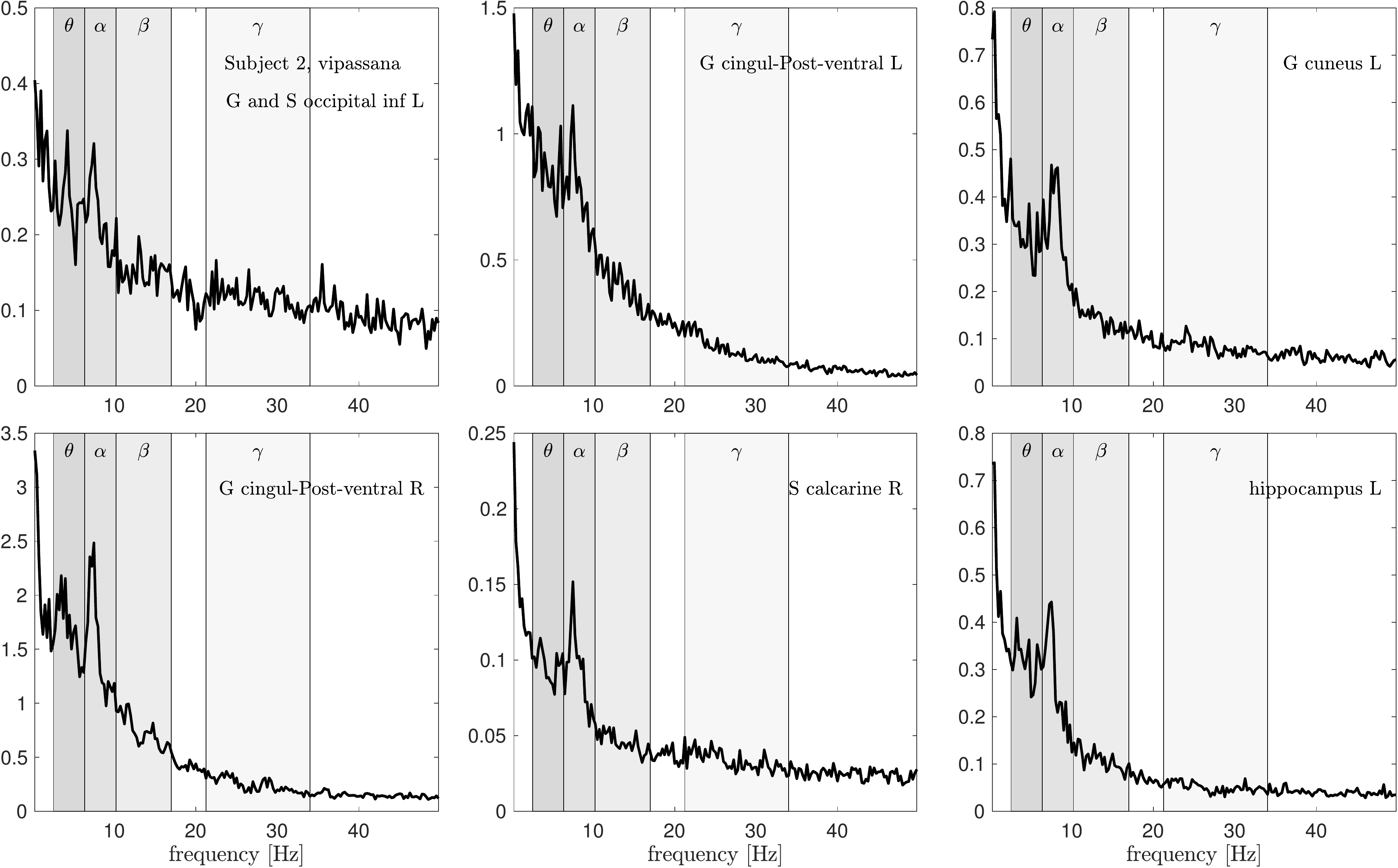}
}
\caption{\label{fig:PSD_3} Six scaled power spectral densities corresponding to open monitoring (Vipassana) meditation state of Subject 1 (two top rows) and Subject 2 (two bottom rows). The scaled PSD values have been scaled by a factor of 100. Observe that each PSD is scaled differently. The peak values for Subject 2 appear at a frequency lower than those of Subject 1, and the frequency bands are adjusted accordingly, see Table 1.}
\end{figure}

Figures~\ref{fig:PSD_1}~--~\ref{fig:PSD_3} show the scaled power spectral densities of six selected brain regions. The densities are calculated using a bootstrap sampling window of length $N=2\,000$, and the number of samples is $p=10\,000$ for each brain state. The regions included in these plots are selected by visual inspection to represent a pronounced spectral spikes, in particular the $\alpha$ spike. The spikes of Subject 1 are more outstanding, while a clear $\beta$ spike is not easy to identify in Subject 2, which may be due to the different quality of the measured data for the two subjects. Importantly, a systematic analysis of the PSDs demonstrates that the $\theta$-, $\alpha$-, and $\beta$-spikes of Subject 2 appear at lower frequencies than those of Subject 1, the ratio of the frequencies being approximately 0.85. Based on this offset, we adjust the frequency bands by the same factor. We do not have enough information about the individuals, or their meditation practices and training to be able to discuss the causes of the observed differences.

\begin{figure}
\centerline{
\includegraphics[width=18.5cm]{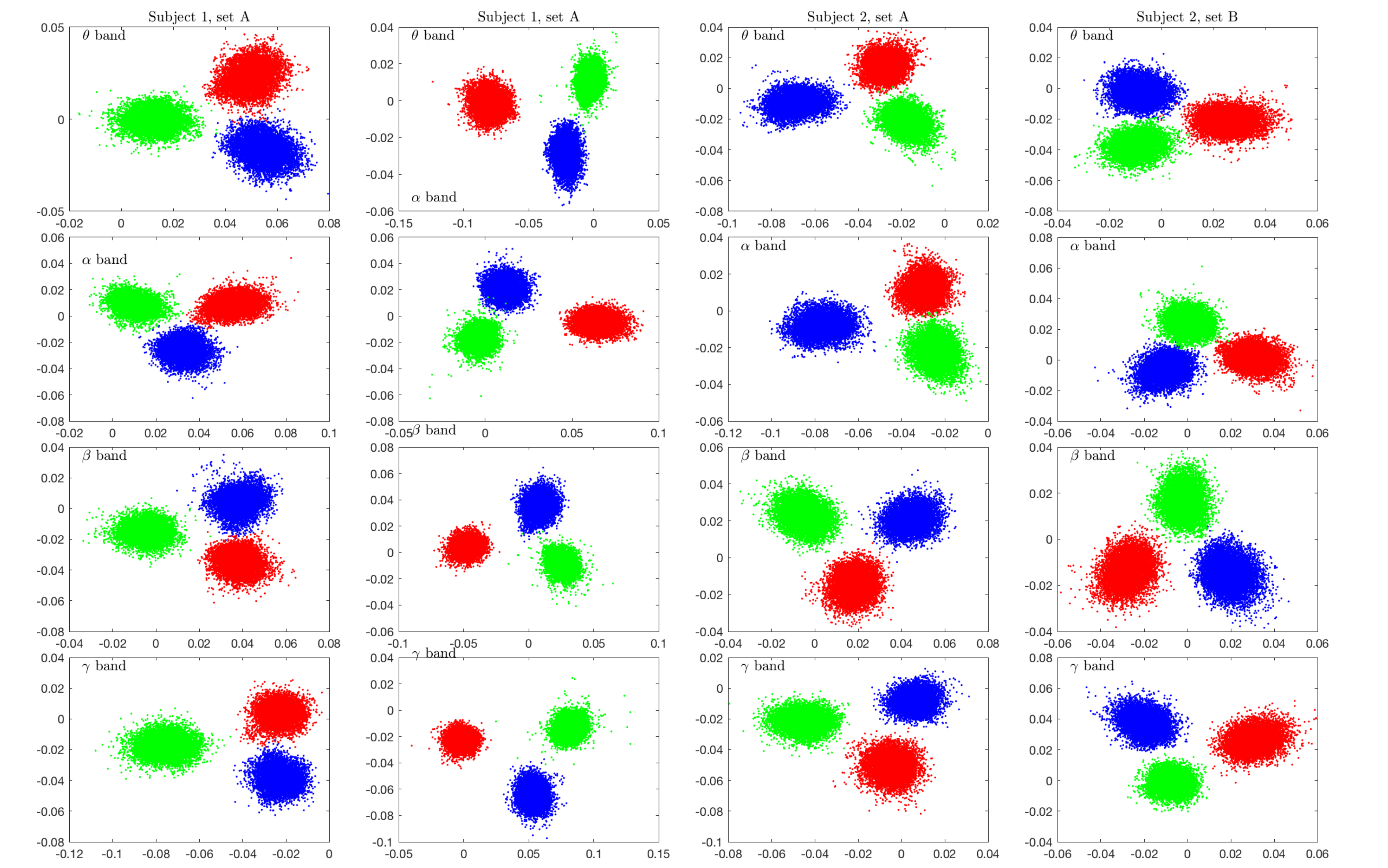}
}
\caption{\label{fig:scatterplots} LDA scatterplots corresponding to the four independent data sets (two sets per subject), and for the four frequency bands. The bootstrap window is $W = 2\,000$. Blue dots correspond to eyes closed resting state, red dots to Samatha, and green dots to Vipassana.}
\end{figure}

Figure~\ref{fig:scatterplots} shows the LDA scatterplots of four data sets, set A and B for both subjects, and for each of the four frequency bands. The DC component has not been included because the arbitrary scaling of the data may affect the separation. Furthermore, the four frequency bands shown here are scaled by the DC component, to remove the possible separating effect due the scaling. We see that the LDA cluster separation is consistent across the different data sets and the frequency bands.

\begin{figure}
\includegraphics[width=18.5cm]{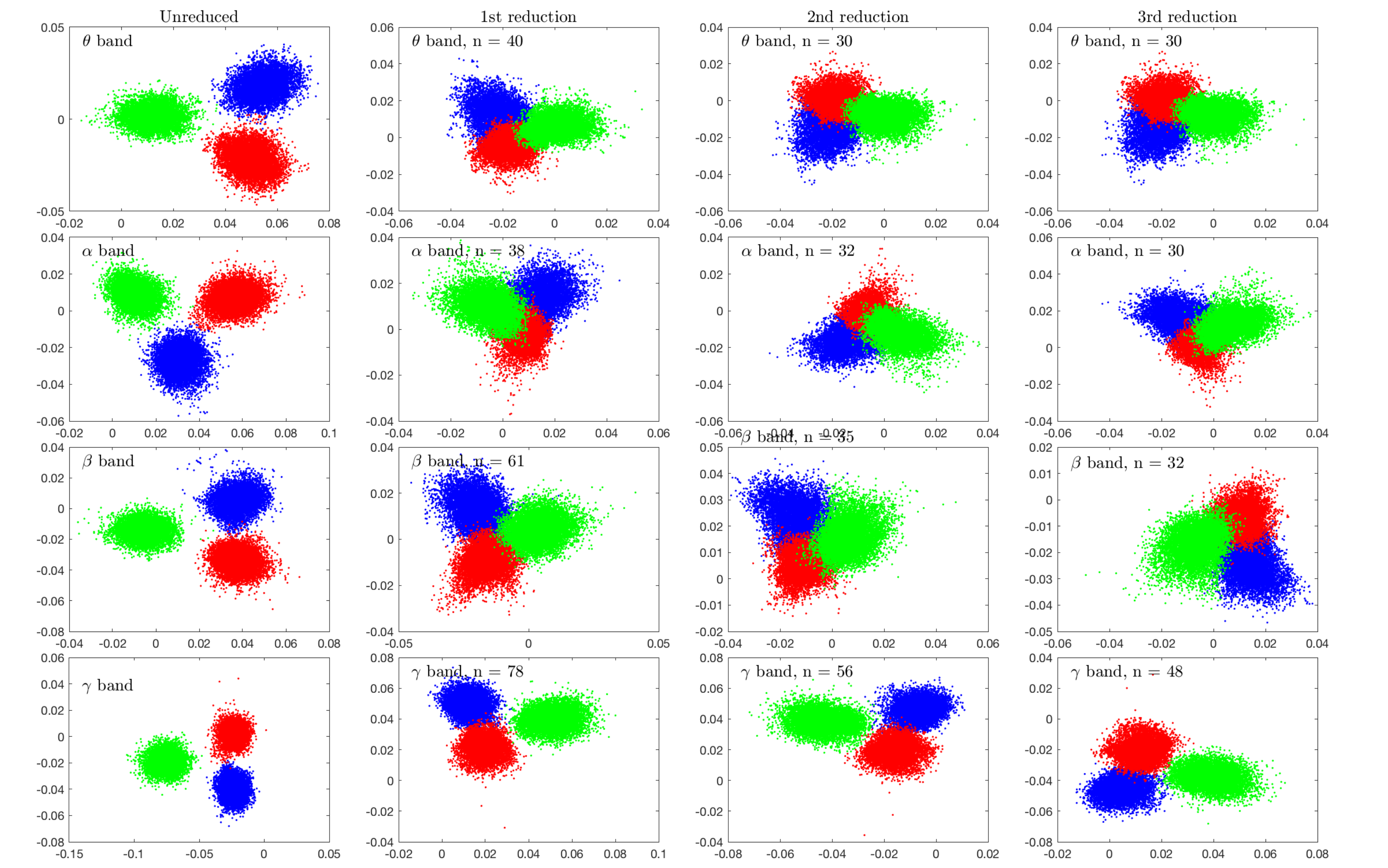}
\caption{\label{fig:sieve} LDA separation combined with LDA-guided model reduction. In the left column, the scatter plots of the LDA components of different spectral bands are shown.  Blue dots correspond to eyes closed resting state, red dots to Samatha, and green dots to Vipassana. In the second column on the left, the brain regions corresponding to components in the LDA separating vectors with less than 10\% of the maximum component are discarded, and the LDA analysis is performed with the reduced model. The process is continued similarly, removing progressively the brain regions with a low component in the separating vector. In each figure, the number $n$ of retained components are indicated.}
\end{figure}

We consider next the LDA separating vectors, and in particular, seek to identify the brain regions that are credited with the most significant role in the LDA separation. To investigate the separating vector significance hypothesis, we start by running the sequential filtering algorithm. Given the data set of three brain states, we first perform the LDA  to produce two separating vectors ${\bf q}^{(1)}$ and ${\bf q}^{(2)}$ for each frequency band.  Next, we discard from the data sets the rows corresponding to those indices $\ell$ that satisfy
\begin{equation}
 |q_{\ell}^{(1)}|<0.1 \max_{1\leq j\leq n} |q_{j}^{(1)}| \quad \mbox{ and } \quad  |q_{\ell}^{(2)}|<0.1 \max_{1\leq j\leq n} |q_{j}^{(2)}|.
\end{equation}
In other words, a brain region is deemed as a significant separator if the amplitude of the corresponding component in at least one of the separating vectors is more than 10\% of the maximum amplitude of the components of that vector. This process leads to different model reduction for each frequency band. We then run the LDA analysis with each of the reduced data matrices, and repeat the process, retaining only a subset of the brain regions of the previous round. In Figure~\ref{fig:sieve}, the scatter plots of the LDA projections  after one, two and three model reductions are shown, and the number of retained brain regions (index $n$ in the figures) is indicated. The plot shows that by discarding brain regions, the separation of the clusters deteriorates, as one would expect, however the separation is not lost even after three reduction steps. While the conclusions are not quantitatively definitive, the result can cautiously be interpreted  to support the separating vector significance hypothesis: If the components with the largest amplitudes would not correspond to separating brain regions, one would expect a rapid loss of the separation.

Assuming that the separating vector significance hypothesis is valid, we then turn to the key question of which brain regions stand out in the separating vectors. To make the analysis statistically more meaningful, we consider all possible mixtures of the three brain states for both subjects. Having two independent data sets (A and B) of each state, we may choose the eyes closed resting state in two ways, the Samatha in two ways, and the Vipassana in two ways, leading to eight different combinations. We perform the LDA for each of the eight combinations, and identify the components whose amplitude is at least 20\% of the corresponding maximum amplitude of that vector. Having thus identified the most prominent components, we compute a tally of the brain regions that were selected at least once.

\begin{figure}
\centerline{\includegraphics[width=7cm]{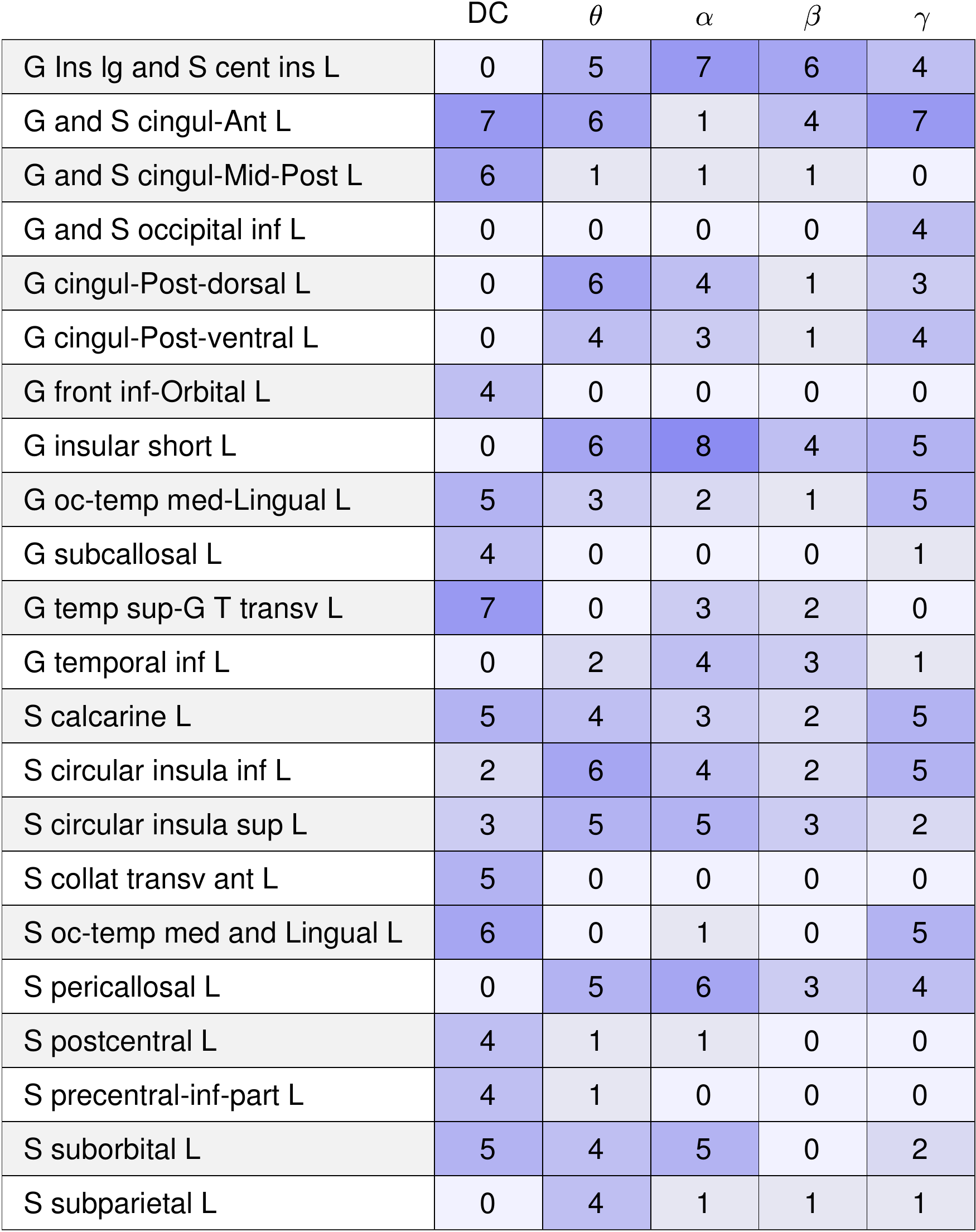} \qquad
\includegraphics[width=7cm]{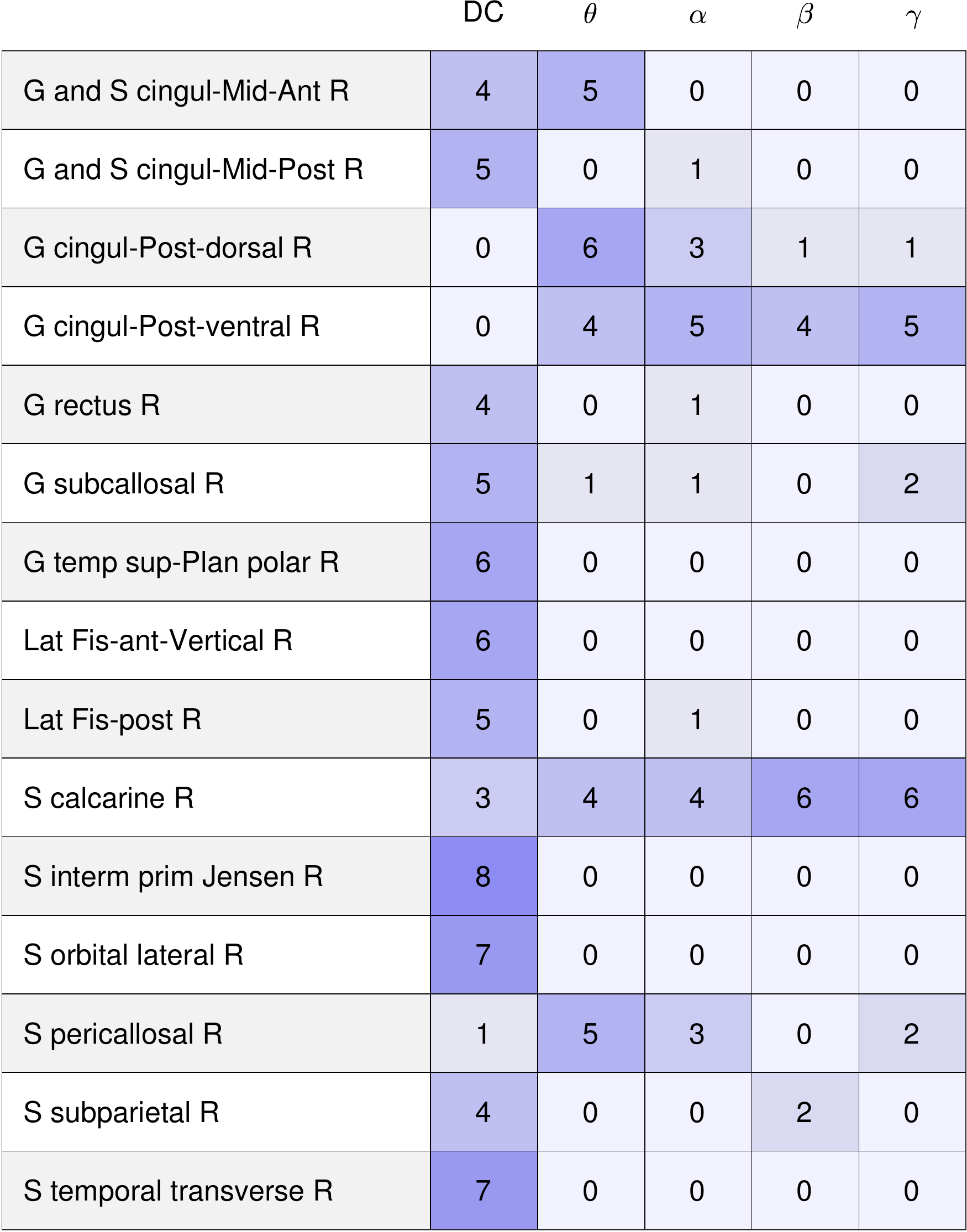}}
\caption{\label{fig:tally cortex M1} Tally of the cortical brain regions of Subject 1 that are identified as having at least 20\% amplitude of the maximum amplitude of the separating LDA vectors. With eight possible combinations of the three meditative states, the tally is at most 8. Only the brain regions that were identified at least once are shown. The shade of the blue is visualizing the tally results.
}
\end{figure}

\begin{figure}
\centerline{\includegraphics[width=7cm]{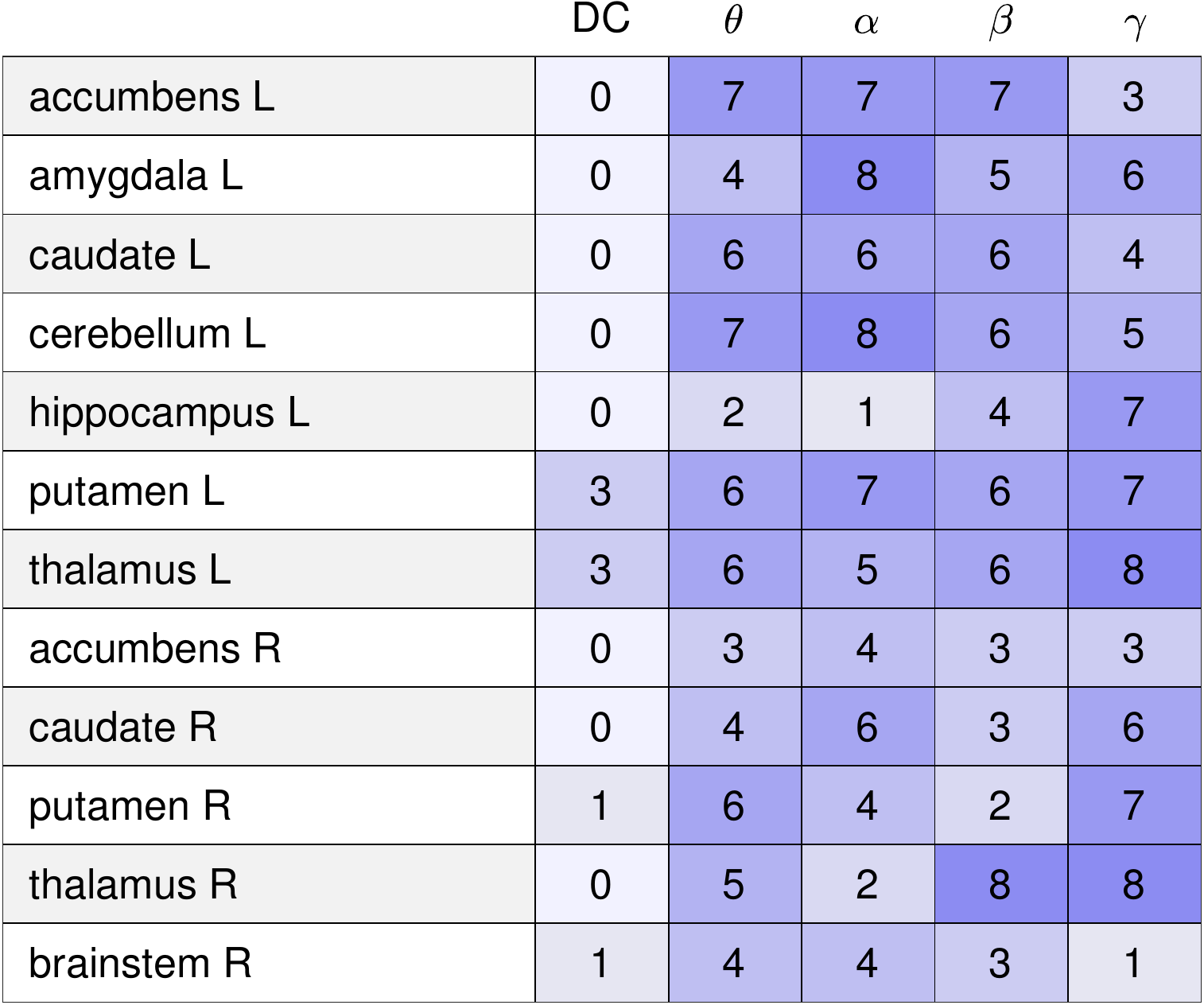}}
\caption{\label{fig:tally struct M1}  Tally of the internal brain regions of Subject 1 that are identified as having at least 20\% amplitude of the maximum amplitude of the separating LDA vectors. With eight possible combinations of the three meditative states, the tally is at most 8. Only the brain regions that were identified at least once are shown. The shade of the blue is visualizing the tally results.
}
\end{figure}

\begin{figure}
\centerline{\includegraphics[width=7cm]{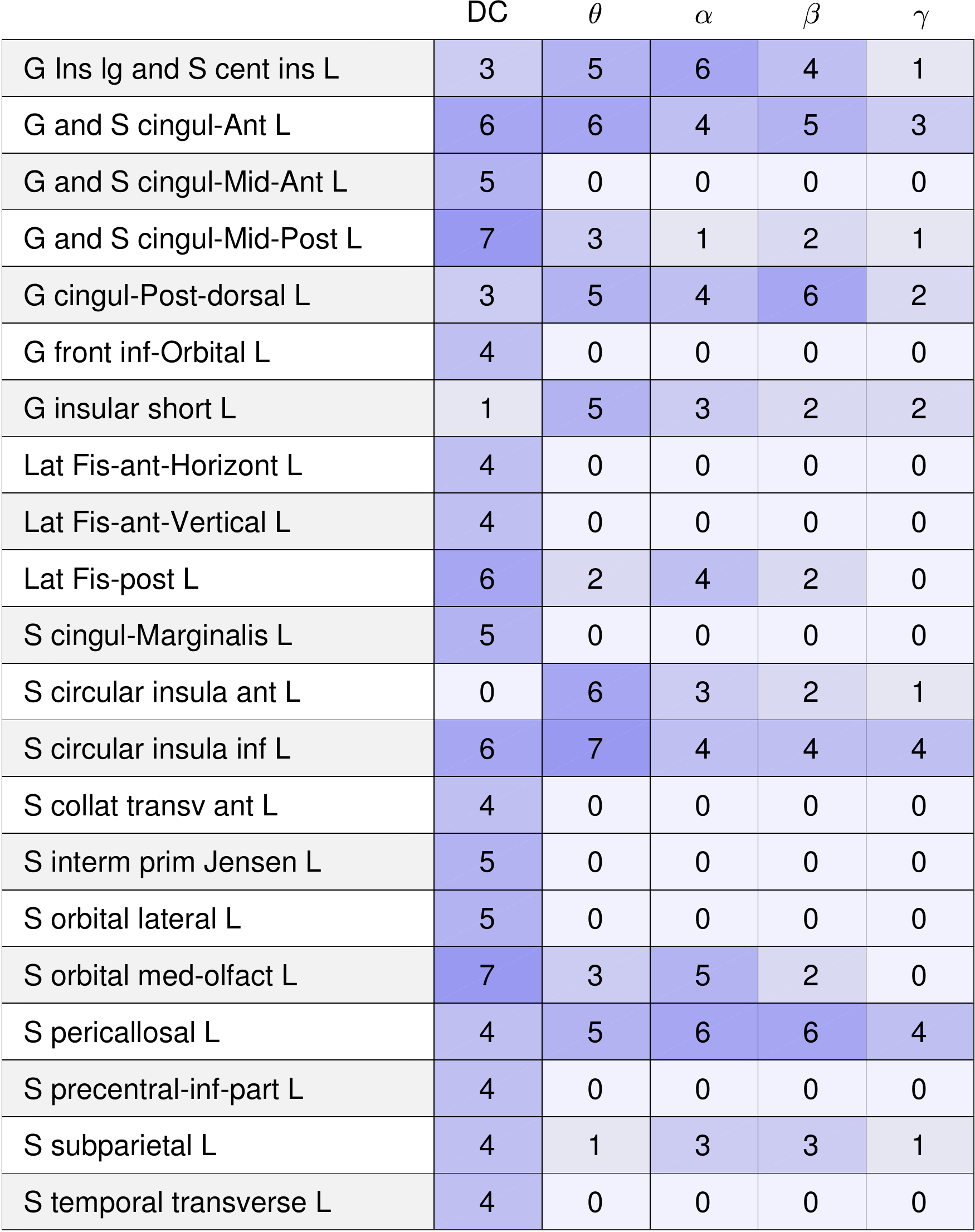} \qquad
\includegraphics[width=7cm]{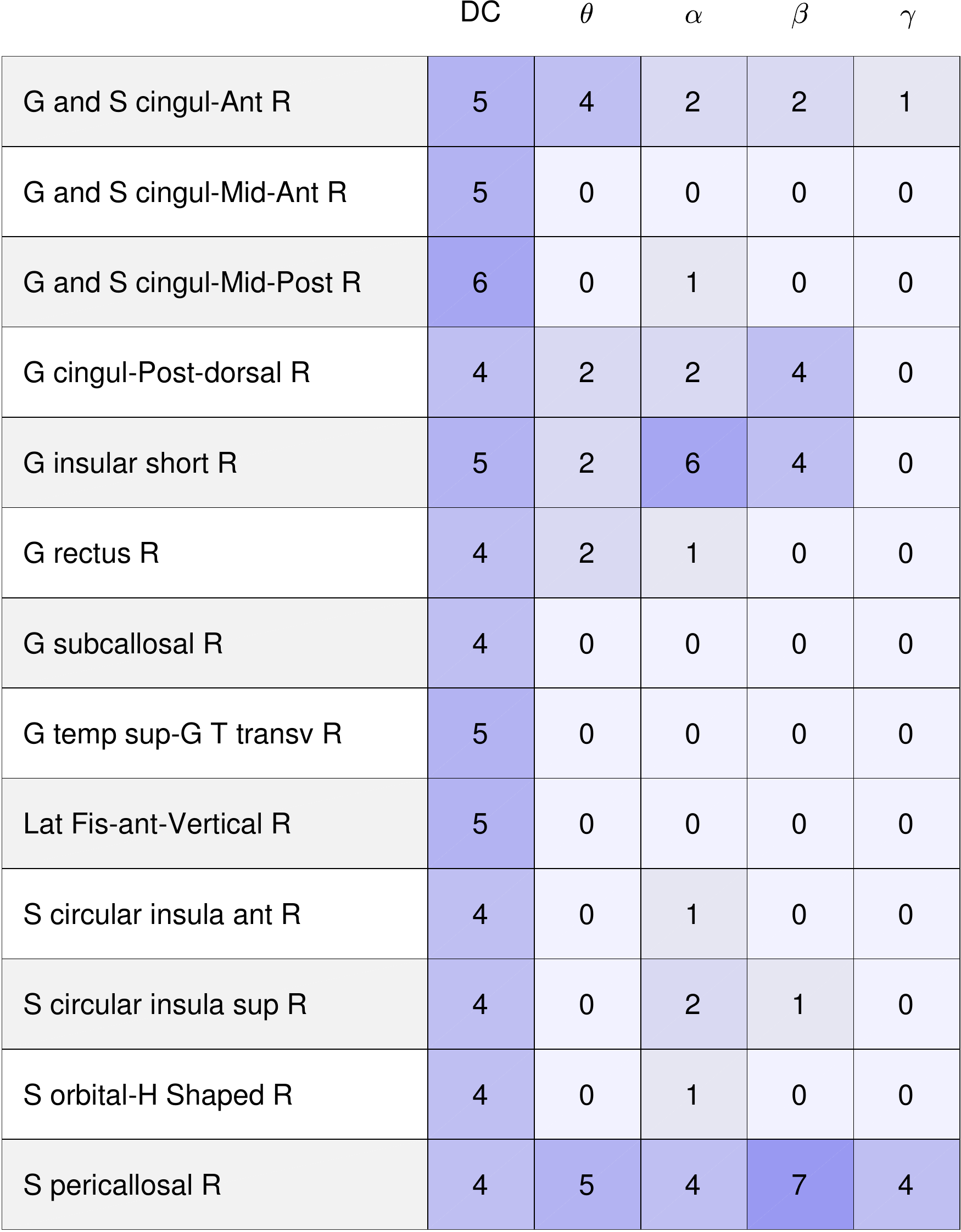}}
\caption{\label{fig:tally cortex M2}  Tally of the cortical brain regions of Subject 2 that are identified as having at least 20\% amplitude of the maximum amplitude of the separating LDA vectors. With eight possible combinations of the three meditative states, the tally is at most 8. Only the brain regions that were identified at least once are shown. The shade of the blue is visualizing the tally results.
}
\end{figure}

\begin{figure}
\centerline{\includegraphics[width=7cm]{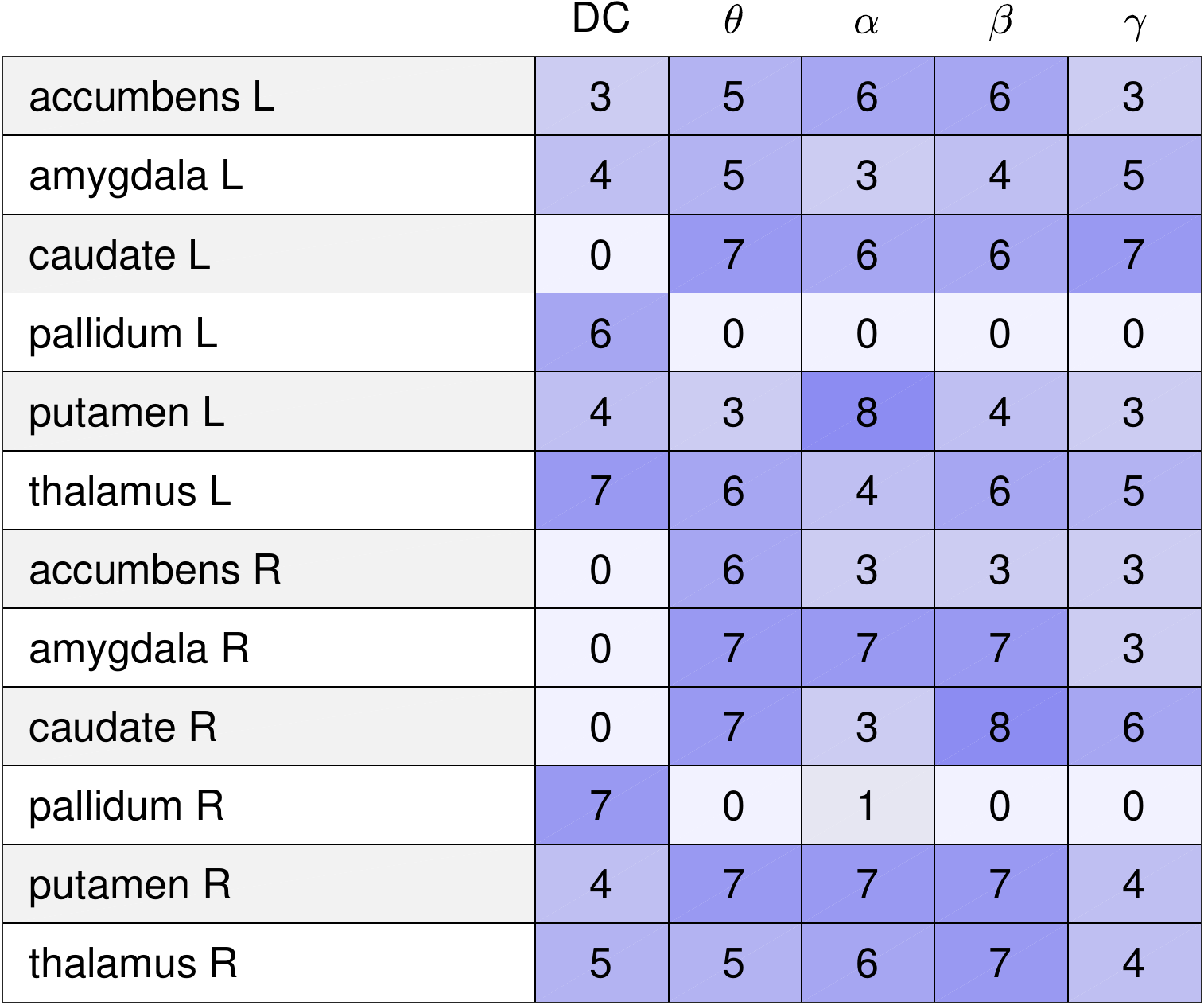}}
\caption{\label{fig:tally struct M2}  Tally of the internal brain regions of Subject 2 that are identified as having at least 20\% amplitude of the maximum amplitude of the separating LDA vectors. With eight possible combinations of the three meditative states, the tally is at most 8. Only the brain regions that were identified at least once are shown. The shade of the blue is visualizing the tally results.
}
\end{figure}

\section{Discussion}

To put the results presented in the previous section into proper context, we briefly review some of the findings reported in the literature. For the most part, the understanding of how meditation affects the brain is not based on functional imaging, and the connection with our findings is therefore not straightforward.

\subsection{Earlier work}

A significant amount of meditation related brain data deals with anatomical changes presumably due to medium to long term (over eight weeks) meditation practice. Consistent anatomical changes in specific brain regions of trained meditators have been reported in the literature.  
For instance, an increase in gray matter density in several brain regions, including inferior temporal gyrus was found by Luders {\em et al.}, who also reported the brain of  long term meditators  to show changes in cortical gyrification \cite{Luders2013}. 

In addition to anatomical changes in the brain, recently there has been a sustained effort to quantify possible changes in brain functions associated with meditation practices, \cite{Fox2016}, in an effort to better understand the mechanism behind meditation and to explore interventional meditation training in clinical settings. 
Dynamic changes in brain functions during meditation are less well understood than morphological ones, partly because of the difficulty of collecting data. While BOLD fMRI may reveal changes in cerebral blood flow, the data collection process is a distracting factor that cannot be neglected. Much less disruptive data collection modalities include EEG and to some extent MEG. MEG data during meditation practice has been previously studied from the point of view of activation patterns and connectivity  \cite{Wong,Aftanas,Yamamoto,Kerr,Marzetti,Lee,Kiebel}, with attempts of  psychological and cognitive interpretation \cite{DorZiderman,Sasaki}, and to understand changes in the sensation of pain \cite{Kakigi}. In the activation studies, the starting point is often the fMRI based functional understanding of activation patterns such as the default mode network (DMN).  A number of neuroimaging studies found that meditation practices induce significative changes in the emotion regulating portion of the brain, in particular on the insula, amygdala and basal ganglia regions, that align with the self-reported state of well being associated with meditation practices.

A central question from the point of view of the analysis in this paper
is whether the effects of meditation, either long term or during the practice, can be characterized in terms of changes in  neural oscillations activated in the different brain regions. Several papers have addressed this question, focusing on different wave bands.  Based on EEG measurements, the $\alpha$ power has been reported higher in Vipassana than during rest mainly at frontal and fronto-central electrodes \cite{Braboszcz}. While $\alpha$ activity is not considered a hallmark of meditative states, it is generally associated with redirection of attention to internal objects, and it has been hypothesized \cite{Posner2018} that an increase in  the $\alpha$ spectral power plays a role in thalamo-cortical sensory transmission, hence relating to functional inhibition, and the suppression of irrelevant input, tantamount to increasing the gating of distracting stimuli.  

Although the significance of increase in power of {$\gamma$ waves (20-100 Hz) is not fully understood, these brain rhythms are correlated with working memory and attention. An increase in $\gamma$ power recorded by EEG has been related, although not in a definitive manner, to ongoing stream and contents of consciousness, and visual representation. Generally, $\gamma$ power over parieto-occipital electrodes is higher in meditators than control, and findings support the hypothesis that training in mindfulness meditation increases the $\gamma$ power in a manner proportional to experience. An EEG-based  study by Lutz el al \cite{Lutz2004}  reported increase in the $\gamma$ band in fronto-lateral and posterior electrodes, while  \cite{Cahn2009} found increased $\gamma$ in parieto-occipital electrode during Vipassana practice.  Another study that compared three styles of meditation found the increase in the higher (60-110 Hz)  $\gamma$ band in electrodes in parieto-occipital area in meditators over control to correlate positively with the experience of the meditators.

The potential for mindfulness meditation as a clinical tool for depression is a topic of active investigation. Yang  et al. \cite{Yang2019, Yang2016}, using resting state fMRI data to investigate the structural and functional changes in the brain of novices following a 40-day training in mindfulness meditation (Vipassana), found a significant thickening of the precuneus region and a decrease in $\alpha$ amplitude.  These changes are suggestive of resting network changes following meditation training, and correlate with reduction in depression scores. A review of mindfulness meditation in substance abuse can be found in  \cite{Zgierska}.

In an effort to correlate changes in brain connectivity and depression states to determine whether it is possible to establish a priori if a subject is likely to benefit from meditation, a recent study  \cite{Doborjeh} applied a  spiking neural network model to EEG measurements in different wave bands before and after a  six week mindfulness meditation training period for non-depressed (ND), depressed before but not after training (D+), and depressed before and after training  (D-) individuals. The findings pointed to an increase in spatio-temporal connectivity over the frontal, centro-parietal and occipito-parietal areas in the ND group after training, with less significant increases for the D+ group, and minimal changes in the D- group. 

In an effort to find coherence between the findings of different groups on which frequency bands that may be altered during a meditation session or following meditation training, Lomas {\em et al.}  \cite{Lomas2015}, after a  systematic review of 56 papers, for a total of 1715 subjects, reported an agreement that $\alpha$ and $\theta$ powers are higher during mindfulness meditation than at rest, while there was no overall consistency in the reported changes in $\beta$, $\gamma$ and $\delta$ power.  Some of the papers, however, found increase in $\theta$, and to a lesser extent in $\alpha$ power in anterior cingulate cortex (ACC) and adjacent prefrontal cortex.

\subsection{Our findings}

In our analysis, we focus on the separating vector components that correspond to the four spectral bands of $\theta$, $\alpha$, $\beta$, and $\gamma$ waves. Although the
overall DC power shows clear separation capabilities, as reported in the results, the risk of having significant separation of the states because of arbitrary scaling of the reconstructions deems the DC separation as potentially less reliable. Our scaling of the periodograms by the DC component makes the frequency band analysis less prone to effects of different preprocessing of each session. To identify the brain regions that play a main role in separating the three brain states, we computed the two LDA separating vectors for samples of time sequences from each state obtained by randomly selecting for each band segments of source reconstructions from data collected during two different sessions for each meditation protocol, for a total of $2^3=8$  possible combinations. The tallies of the cortical brain regions of Subject 1 that are identified as having at least 20\% amplitude of the maximum amplitude of the separating LDA vectors are reported in Figure~\ref{fig:tally cortex M1}, while Figure~\ref{fig:tally struct M1} reports the results relative to the internal brain structures for Subject 1. With eight possible combinations of the three meditative states, the tally is at most 8. Only the brain regions that were identified at least once are shown. The analogous results for Subject 2 are summarized in Figure~\ref{fig:tally cortex M2} (cortical regions) and Figure~\ref{fig:tally struct M2} (internal structures).

In the following, we highlight some of the brain regions that appear to play a significant role in the separation of the activities, and present a short review of previous literature that supports the findings.

\subsubsection{Anterior and posterior cingulate cortex (ACC and PCC)}

The cortical regions consistently associated with the support of the
separating vectors for the three brain states are located in the proximity of  the internal structures, with the left hemisphere more conspicously
represented than the right one. The separator vectors for both subjects show a large component in all four spectral bands, corresponding to the anterior
part of the cingulate gyrus and sulcus (ACC), posterior-dorsal part of the cingulate gyrus (dPCC) and pericallosal sulcus in both hemisphere;
see: Figure~\ref{fig:ACC} for the main discriminant regions\footnote{Figure produced by open source visualization software Visbrain \cite{visbrain}}.

\begin{figure}
\centerline{\includegraphics[width=14cm]{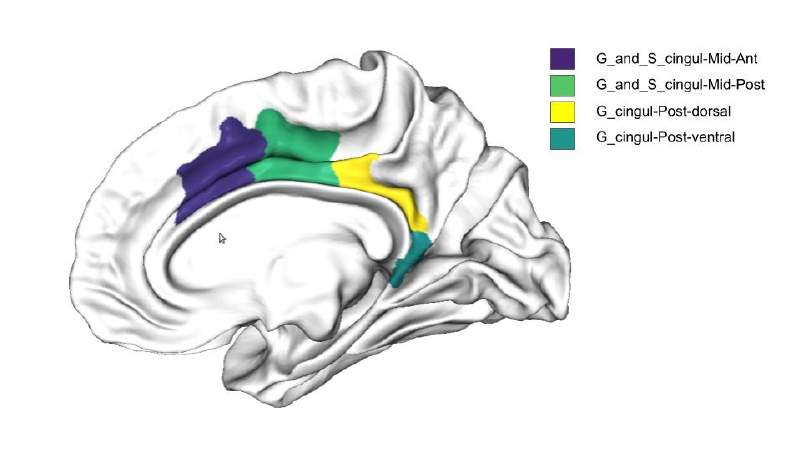}}
\caption{\label{fig:ACC} The figure shows the ROIs in the anterior
and posterior cingulate cortex (right hemisphere) that are found to play a significant role
in the separation of meditative and resting brain states.}
\end{figure}

The cingulate cortex is an integral part of the limbic system, which is involved with the formation and processing of emotions as well as learning and memory processes. Changes in the posterior cingulate cortex in connection with mindfulness meditation have been the topic of several studies, some concentrating on its connectivity as part of the DMN, other reporting lower PCC activity in trained meditators than in novices. In light of the observation that the non self-referential nature of mindfulness meditation is likely to decrease PCC activity \cite{QinNorthoff, BreczynskiLewis}, it is not surprising that this BR plays a role in the separation, although our investigation does not specify whether
the open monitoring meditation (Vipassana, or mindfulness meditation) increases or decreases  activation in PCC. The ACC, also appearing as one of the BRs playing a role in separating the three brain states, is generally believed to contribute to cognitive control and emotional regulation. 

Increased gray matter density in left hippocampus, PCC,  temporo-parietal junction, and cerebellum, all regions involved in learning, memory processes, emotion regulation self-referential and perspective processes, has been reported \cite{Holzel2011} after eight weeks of mindfulness meditation training.
Professional meditation activity is a long and demanding path, therefore a question that has received attention in the literature, in the context of using mindfulness meditation as an interventional clinical tool, is whether a short-term meditation training can change the brain structure and functions \cite{Kozaka2018}. Higher fractional anisotropy around the ACC, measured via diffusion tensor imaging after two weeks of mindful meditation, reported in \cite{TangMa2009}, is considered as a proxy for increased white matter pathways. Changes in myelination and axon density after four weeks of mindful meditation have been reported in \cite{Tang2010}. 

Dunlop et al \cite{Dunlop}, who examined mindfulness meditation in connection with treatment of major depression disorder, reported functional connectivity of sub-callousal cingulate cortex (SCC) with left anterior ventrolateral prefrontal cortex/insular dorsal midbrain, and left ventromedial prefrontal cortex. Major depression is associated with relative hyperactivity in limbic brain regions, including amygdala, insula and SCC, and hypoactivity in dorsolateral prefrontal cortex. Moreover, elevated metabolism in SCC appeared to be a predictor of poor outcome of antidepressants. 

A number of functional studies focus on changes in the DMN, which is known to deactivate during the performance of tasks.   A review of the literature on functional and structural neuroimaging studies of the neural processes associated with mindfulness meditation \cite{Marchand} provides compelling evidence of impact on the medial cortex  and associated DMN, insula and amygdala, and some evidence of effects on basal ganglia and latero-frontal regions. The role of meditation in the regulation of brain networks continues to be actively studied. Changes in the DMN, and in particular on the PCC connectivity, associated with mindfulness meditation practice have been addressed in several studies, motivated by the observation that higher DMN activation correlates with attention lapse and anxiety. Reduction in DMN activity during the practice of mindfulness meditation was first described in \cite{Brewer2011}. Decrease in PCC activity during mindfulness meditation is quite plausible, as this is a non self-referential task, and some studies \cite{QinNorthoff, BreczynskiLewis} show that trained monks have lower PCC activity than novice meditators. Furthermore, while investigating a possible role of mindfulness meditation in smoke cessation and other behavioral interventions, Brewer {\em et al.} \cite{Brewer2011b, Brewer2013} identified the DMN, and in particular the PCC, as some of the regions primarily affected by meditation. The analysis of MEG recording in a cohort of experienced meditators \cite{Marzetti} found the coupling in the $\alpha$ frequency band of the PCC with different brain regions to change according to the type of meditation practice. A difference in the level of $\alpha$ activity between experienced and novice meditators was reported in \cite{Wong}.

Tang et al  \cite{Tang2019} found increase in $\theta$ power in frontal midline (FM) EEG electrode following mindfulness meditation and hypothesized that the increase may result in a proliferation of oligodendrocytes, hence increase myelination and in turn improve the connectivity between ACC and the limbic areas; see also \cite{PosnerTang2014, Posner2018}. Earlier on, an increased $\theta$ activity was found to be positively correlated with glucose metabolism in the ACC \cite{Pizzagalli}. 
EEG dynamics in Vipassana meditators \cite{Kakumanu} shows increase in $\delta$ (1-4Hz) and low $\gamma$ (30-40Hz) power at baseline, and increase in $\alpha$ and low $\gamma$  power during mindfulness practice. 

\subsubsection{Insular cortex}
  
The insular cortex (left for Subject 1, and both sides for Subject 2, see Figure~\ref{fig:insula} for the main discriminant regions\footnote{Figure produced by open source visualization software visbrain \cite{visbrain}}) seems to have a significant prominence in several of the frequency bands. The insular cortex is credited with playing a significant role in a range of processes, including feelings and emotions, bodily- and self-awareness, decision making, sensory processing, and social functions such as empathy \cite{Gogolla}. The central role of insular cortex in meditation practice has been implicated in \cite{Luders2012} where an increase in cortical gyrification in meditators was observed. A cortical thickening of the insular cortex of long term meditators, observed in \cite{Engen}, was similarly concluded to point towards a central role of this brain region in meditation. The importance of insular cortex in connection with meditation was highlighted in \cite{Dunlop} also.
In  \cite{Mooneyham2017}, the authors report increases in the left hemisphere posterior insula and in its functional connectivity with middle and superior left temporal gyrus, and right ventrolateral prefrontal cortex after a six weeks' training bootcamp in mindfulness meditation. These brain regions comprise a structurally connected network involved in early auditory perception, allocation of attention to changes in sound and analysis of sound fluctuations.  In \cite{Kirkpatrick}, meditators were found to have increased connectivity between prefrontal cortex (PFC) network and posterior insula.

\begin{figure}
\centerline{\includegraphics[width=14cm]{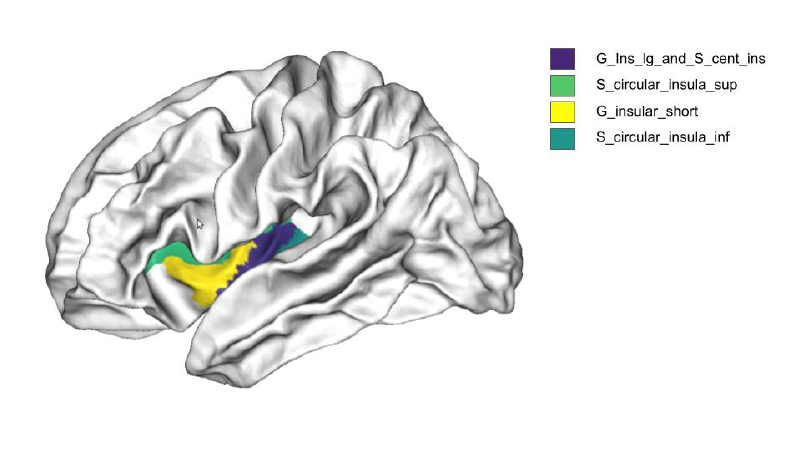}}
\caption{\label{fig:insula} The figure shows the ROIs in left insular cortex that are found to play a significant role in the separation of meditative and resting brain states.}
\end{figure}

\subsubsection{Internal structures}

In spite of their relative small volume and distance from the sensors, the internal structures have the lion's share  among the attributes most relevant
for separating the three brain states. More specifically, the left and right accumbens, caudate and putamen nuclei, together with left and right thalamus,
and left amygdala are significantly represented in the support of the separating directions for both subjects (see Figure~\ref{fig:deep}\footnote{Figure produced by Brainstorm software \cite{Tadel}} ). The putative roles of these regions in the regulation of activities and emotions is in line with the reported different mental states associated with the three protocols.  The nucleus accumbens and  caudate are among the internal structures consistently represented in the vectors separating the three different states for both subjects in all frequency bands. The accumbens, a region in the ventral striatum, has been extensively studied over the last half century, and  because of its involvement with the reward circuit is often referred to as  pleasure center of the brain. In a recent comprehensive review of the role of the nucleus accumbens \cite{Floresco2015nucleus}, this brain region has been characterized as an aid to obtain motivational relevant goals by promoting behaviors leading to their attainment.  This aligns well with the state of well being usually associated with Vipassana, and the steps needed to reach it.

Another striatum region,  with a significant role in separating brain states for both subjects, is the caudate nucleus, also related to the reward system and part of the cortico-ganglia-thalamic group. The caudate nucleus,  typically associated with a number of clinical conditions related to movement, including Parkinson and Huntington disease, has a significant role in cognitive functions associated with learning and memory. A retrospective review of the literature on the different roles of the caudate and accumbens nuclei \cite{Grahn} concluded that the caudate contributes to learning and memory by overseeing processing that are fundamental to all tasks involved in goal-directed actions. In light of the strong goal-directed component in  Samatha and Vipassana meditation, it is not surprising that the level of activation of the caudate nucleus acts is one of the factors implicitly used by LDA to cluster and separate the different states. 

The activity in all frequency bands in the putamen nucleus which together with the caudate nucleus forms the dorsal striatum, the region over and to the side of the limbic system, also consistently appears among the attributes characterizing brain changes during meditation. The putamen, whose anatomical structure is very similar to that of the caudate, while mostly associated with preparation of limb movements, has been shown to have a role in category learning \cite{Ell} also. 

The amygdalae, also part of the basal ganglia, are credited with a role in emotions and behavior, especially in response to fear and anxiety. Electrical stimulation of the right amygdala has been shown to elicit negative emotions, in particular anxiety and sadness, while  electrical stimulation of the left amygdala has produced either positive or negative emotions. The consistent presence of the left amygdala among the separating regions suggests changes in its activation at all frequency band levels during meditation practices. This is in line with literature reporting structural changes of the amygdala in relation to stress reduction \cite{Holzel2010}, and reduction of the volume of the right amygdala related to meditation and yoga practices \cite{Gotink}. Effects of mindfulness meditation on the amygdala are reported also in \cite{Marchand}. The presence of the left and right thalamus among the BRs represented in the separating vectors for all frequency bands reinforces the hypothesis that meditation induced changes occur mostly in deeper brain regions. Anatomical investigations found  increase in gray matter volume of left hippocampus, thalamus and caudate among yoga practitioners versus control \cite{Gothe}.  A recent MRI-based study \cite{Kral} found the impact of meditation practice in the form of decreased reactivity of the amygdala in response to positive pictures and increased connectivity of the amygdala with ventromedial prefrontal cortex, a region implicated in emotion regulation and in the processing of self-referential stimuli \cite{Northoff}. 

 A retrospective study \cite{Desai} to determine physiological and structural changes of yoga and meditation on brain waves reported an increase in gray matter and the amygdala, and frontal cortex activation, raising the possibility of using these practices in clinical treatment and to promote successful aging \cite{Sperduti2017}. Other studies corroborate the increase in gray matter volume of left hippocampus, thalamus and caudate among yoga practitioners versus controls \cite{Gothe}. Changes in hippocampal anatomy in long-term meditators, in the form of a larger volume, where reported in  \cite{Luders2012}. 
 
 The MRI-bases study on the role of basal ganglia in  \cite{Gard2015} takes a hypothesis-free approach, by looking at all brain regions instead of limiting the attention to subnetworks. The data, consisting of time series measurements at resting states of  yoga practitioners, Vipassana practitioners and control subjects, found that, in Vipassana and yoga practitioners, the caudate is more connected to other brain regions than in the control group, in line with the increased caudate activity.
The fMRI-based study in \cite{Baerentsen} found increased putamen activity at the onset of meditation and increased caudate activity during sustained meditation; another study \cite{Sperduti2012} also proposes a model of meditative state involving putamen and caudate. The involvement of caudate in mindfulness meditation confirms the related caudal connectivity to cognition, emotion, action and perception \cite{Robinson}, and is in line with goal-oriented learning being meditated by caudate, and habitual learning by putamen \cite{Braunlich}, and stress having been defined as a shift from goal orient to habitual behavior \cite{SchwabeWolf} .

\begin{figure}
\centerline{\includegraphics[width=16cm]{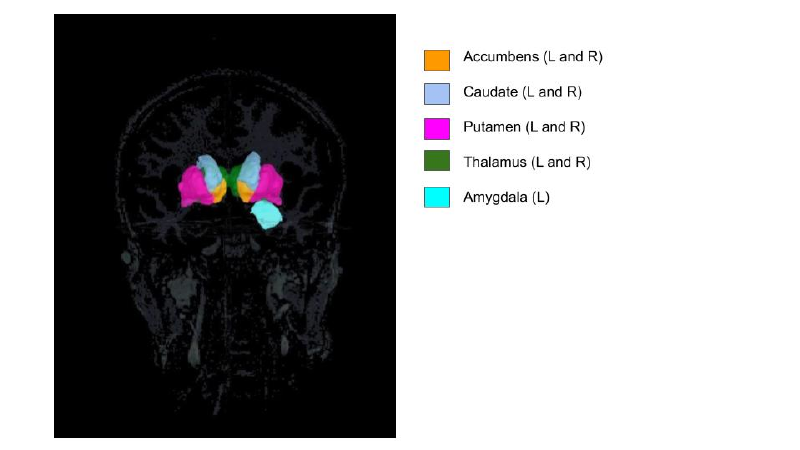}}
\caption{\label{fig:deep}The figure shows the interior structures in both hemispheres that play a significant role in the separation of activities.}
\end{figure}

\section{Conclusions}

The present contribution addresses the problem of whether it is possible to quantify dynamic changes in brain activity during meditation practices versus rest based on the MEG data collected during meditation sessions. The work focuses on four frequency bands, and uses data analysis tools to identify brain regions in which the activity undergoes changes significant enough to give rise to an identifiable separation of the bootstrapped periodograms computed from the spatially resolved sequence of brain activity. The results indicate that the internal structures, constituting the core of the limbic system, play a fundamental role in the separation. In addition, certain cortical areas, most notably insula and cingulate cortex, stand out as possible areas in which the activity differs from resting state activity during the practice. The findings are in line with existing literature that for the most part focuses on anatomo-morphological changes in the brain trained by meditation practice. 

\section*{Acknowledgements} The work of Daniela Calvetti was partly supported by the NSF grants DMS-1522334 and DMS-11951446, and of Erkki Somersalo  by the NSF grant DMS-1714617.
This work was partially done during the visit of DC at University of Rome ``La Sapienza'' (Visiting Researcher/Professor Grant 2018).\\ 
The authors thank Vittorio Pizzella for providing the MEG recordings used in this work.}

\pagebreak

\end{document}